\documentclass[prx, notitlepage, twocolumn,nofootinbib,superscriptaddress]{revtex4-1}
\usepackage[colorlinks=true,citecolor=blue,linkcolor=blue]{hyperref}
\usepackage{graphicx}
\usepackage{color}
\usepackage{dcolumn}
\usepackage{bm}
\usepackage{mathrsfs}
\usepackage{amsmath}
\usepackage{amssymb}
\usepackage{amsthm}
\usepackage{amsfonts}
\usepackage{enumerate}
\usepackage{latexsym}
\usepackage{hyperref}
\usepackage{centernot}
\usepackage{multirow}
\usepackage{longtable}

\begin{document}

\title{Determining the range of magnetic interactions from the
relations between magnon eigenvalues at high-symmetry $k$ points}

\begin{abstract}
Magnetic exchange interactions (MEIs) define networks of coupled magnetic
moments and lead to a surprisingly rich variety of their magnetic
properties. Typically MEIs can be estimated by fitting experimental results.
But how many MEIs need to be included in the fitting process for a material
is not clear a priori, which limits the quality of results obtained by these
conventional methods. In this paper, based on linear spin-wave theory but
without performing matrix diagonalization, we show that for a general
quadratic spin Hamiltonian, there is a simple relation between the Fourier
transform of MEIs and the sum of square of magnon energies (SSME). We
further show that according to the real-space distance range within which
MEIs are considered relevant, one can obtain the corresponding relationships
between SSME in momentum space. We also develop a theoretical tool for
tabulating the rule about SSME. By directly utilizing these characteristics
and the experimental magnon energies at only a few high-symmetry $k$ points
in the Brillouin zone, one can obtain strong constraints about the range of
exchange path beyond which MEIs can be safely neglected. Our methodology is
also general applicable for other Hamiltonian with quadratic Fermi or Boson
operators.
\end{abstract}

\date{\today }
\author{Di Wang}
\affiliation{National Laboratory of Solid State Microstructures and School of Physics,
Nanjing University, Nanjing 210093, China}
\affiliation{Collaborative Innovation Center of Advanced Microstructures, Nanjing
University, Nanjing 210093, China}
\author{Jihai Yu}
\affiliation{National Laboratory of Solid State Microstructures and School of Physics,
Nanjing University, Nanjing 210093, China}
\affiliation{Collaborative Innovation Center of Advanced Microstructures, Nanjing
University, Nanjing 210093, China}
\author{Feng Tang}
\affiliation{National Laboratory of Solid State Microstructures and School of Physics,
Nanjing University, Nanjing 210093, China}
\affiliation{Collaborative Innovation Center of Advanced Microstructures, Nanjing
University, Nanjing 210093, China}
\author{Yuan Li}
\affiliation{International Center for Quantum Materials, School of Physics, Peking
University, Beijing 100871, China}
\affiliation{Collaborative Innovation Center of Quantum Matter, Beijing 100871, China}
\author{Xiangang Wan}
\thanks{The corresponding author: xgwan@nju.edu.cn.}
\affiliation{National Laboratory of Solid State Microstructures and School of Physics,
Nanjing University, Nanjing 210093, China}
\affiliation{Collaborative Innovation Center of Advanced Microstructures, Nanjing
University, Nanjing 210093, China}
\maketitle

\section{Introduction}

As one of the oldest scientific topics, magnetism is still of great interest
\cite{book-1,book-2,book-3,Lichtenstein-book}. Magnetic materials had
already been widely used in electromechanical and electronic devices, and
its applications in information technology are also continuously growing
\cite{book-1,book-2,book-3}. Especially magnons, as the quanta of spin
waves, have received more and more research attention over the past few
decades \cite{spinwavebook,spin-wave-1}. As an elementary excitations of
magnetic systems, magnons became an interesting platform for the study of
general wave dynamics \cite{kosevich1990magnetic,fogedby1980solitons},
Bose-Einstein condensation of magnon \cite%
{giamarchi2008bose,nikuni2000bose,demokritov2006bose} and so on. In
addition, with the development of topological physics in the electron
system, topology in magnon spectrum has also attracted significant interests
\cite{onose2010observation,chisnell2015topological,kondo2019z}, including
topological magnon insulators \cite{topomagnonti-1,topomagnonti-2}, magnonic
Dirac semimetals \cite%
{topomagnondirac-1,topomagnondirac-2,topomagnon-new,CTO-ref2,CTO-ref3} and
Weyl semimetals \cite{topomagnon-1,topomagnon-2,topomagnon-3}. Besides
fundamental research, magnons have also attracted great attention for
applications of information transport and processing \cite%
{serga2010yig,kruglyak2010magnonics,chumak2015magnon,Magnonics-4,Magnonics-5}%
. Analogous to spintronics, the application of magnon are connected with the
ability to carry, transport and process information. Potentially, the spins
can be manipulated without current, thereby overcoming an important
fundamental limitation of conventional electronic devices, the dissipation
of energy due to Ohmic losses. Magnon spintronics is therefore an emerging
field of modern magnetism, which has spurred significant advances towards
computing application recently and is believed to deliver a number of
breakthrough developments in the future \cite%
{serga2010yig,kruglyak2010magnonics,chumak2015magnon,Magnonics-4,Magnonics-5}%
. In order to quantitatively understand the rich phenomenon and wide
applications in this highly interdisciplinary field, a microscopic magnetic
model with proper parameters becomes extremely important.

Magnetic properties can be typically described by a quadratic spin
Hamiltonian $H=\sum_{i,j}\mathbf{S}_{i}\cdot \mathbf{J}_{ij}\cdot \mathbf{S}%
_{j}=\sum_{i,j}J_{ij}^{\alpha \beta }S_{i}^{\alpha }S_{j}^{\beta }$ , where $\mathbf{J}_{ij}$ represents the magnetic exchange interaction (MEI) between the spin
at $i$ site $\mathbf{S}_{i}$ and spin at $j$ site $\mathbf{S}_{j}$ as shown
in the following Eq. (\ref{1}). The sum should take over all possible
exchange paths with sizable MEIs. However, it turns out that extracting
quantitative $J_{ij}$ is a highly non-trivial task. By choosing the set of
parameters that best fit the experimental results, such as
temperature-dependent magnetization, magnetic susceptibility $\chi (T)$,
magnetic excitation spectra $\omega (q)$ etc, one basically can obtain MEIs $%
J$'s \cite{book-1,book-2,book-3,Lichtenstein-book}. It is well known that
the $J_{ij}$ usually decreases rapidly with increasing of $R_{ij}$, the
distance between magnetic moment at $i$ and $j$ sites, and the $J$'s with
sufficient distance are believed to be negligible. Thus only a number of $J$%
's within a cut-off range $R_{cut}$ are needed to be considered. However, a
priori knowledge about $R_{cut}$ is unknown, while the number of MEI used to
fit the experimental data obviously affect the obtained $J$'s. This leads to
the arbitrariness of fitting approach, consequently affected the accuracy of
the estimated MEIs, and currently unambiguous fitting is basically
impossible. For example, very similar inelastic neutron scattering (INS)
experimental results can be fitted by considerably different MEI parameters
\cite{CTO-ref2,CTO-ref3}.

In addition to the above approach, theoretical calculations had also been
used to evaluate the exchange interaction parameters \cite%
{xiang2013magnetic,Jref1,Bruno,ourJ1,KKR-J-1,KKR-J-2,frozen-magnon-1,frozen-magnon-2,Paddison}%
. A popular numerical method is to calculate the total energies of more than
$N$ magnetic configurations and map them using a spin Hamiltonian to extract
$N$ MEIs \cite{xiang2013magnetic}. Unfortunately, this method also need to
assume a cut-off range $R_{cut}$, which again leads to the arbitrariness
about the calculated MEIs. An alternative method is based on combining
magnetic force theorem and linear-response approach \cite%
{Jref1,Bruno,ourJ1,KKR-J-1,KKR-J-2}. Working in the momentum space, this
method indeed does not suffer the problem about $R_{cut}$. However, the
Coulomb interaction which had been incorporated by the parameter $U$ in first-principles calculations, usually play important role in magnetic systems
\cite{LDA+U,LDA+DMFT}. Thus these theoretical MEIs strongly depend on the choice of the parameter $U$ \cite{LDA+U,LDA+DMFT}.

Symmetry imposes constraints about the magnetic model, and one can also use
symmetry to check if two exchange paths with the same bond length have the
same MEI. Unfortunately, this powerful theoretical method cannot provide
any clue about $R_{cut}$. The general features, such as sum rule for the
spectral weight of the spin correlation function \cite{sum-rule} which requires accurate cross-section measurements over the entire Brillouin zone (BZ), also cannot predict the variation of MEIs over distance.
Thus to explore possible
$R_{cut}$-related generic rules is a very important problem. Certain
important subjects on magnetism, such as quantum spin liquids arising from
exactly solvable models \cite{Kitaev}, novel properties from geometrically
frustrated magnet \cite{pyrochlore}, etc., explicitly requires small $%
R_{cut} $, hence gaining a wealth of knowledge for $R_{cut}$ in a large set
of known magnetic materials will also be empirically useful for assessing
the relevance of such models.

In this work, based on linear spin-wave theory (LSWT), we find that for a
general quadratic spin Hamiltonian, the sum of square of magnon energies
(SSME) at arbitrary $k$ point in BZ can be directly obtained
by the Fourier transform of MEIs, consequently one can easily calculate SSME
at arbitrary $k$ point in BZ without diagonalization. Thus,
different from conventional symmetry analysis which groups the magnon energies into symmetry-related $k$ points, our method produces
different relationships between the SSME at different high-symmetry
$k$ points subjected to different $R_{cut}$. Thus, using the magnon energies
at only several high-symmetry $k$ points, which can be measured by inelastic
neutron scattering accurately \cite%
{book-1,book-2,Lichtenstein-book,book-3,spinwavebook,spin-wave-1}, one can
unambiguously assert up to which neighbor the MEIs becomes negligible. To demonstrate how our algorithm works,
we show an example for Heisenberg model with ferromagnetic (FM) configuration and give the
discussion about general cases with DM interaction, single-ion anisotropy
(SIA) as well as non-collinear magnetic ordering. Instead of exhaustedly
listing the SSME relationships for all magnetic space group (MSG), we
provide a code in the Supplemental Materials (SM). With the basic
information about a magnetic material (i.e. space group, the positions and
magnetic moments orientations of the magnetic ions), the code will deliver
corresponding SSME relationships according to the input $R_{cut}$. Thus
through simply checking up to which $R_{cut}$, the experimental SSME start
to deviate from the obtained theoretical relationships, one can determine
the real-space range of sizable MEIs. Our method can be easily extended to
other Hamiltonian with quadratic Fermi or boson operators, thus is useful
for the characteristics of the electronic band structure, phonon spectrum,
etc.

\section{Method}

Usually the magnetic properties of crystal materials can be well described
by a general pairwise magnetic model \cite%
{book-1,book-2,book-3,Lichtenstein-book,spinwavebook,spin-wave-1}

\begin{eqnarray}
H &=&\sum_{l,n,\alpha ,l^{\prime },n^{\prime },\beta }J_{\mathbf{R}_{l}+%
\boldsymbol{\tau }_{n},\mathbf{R}_{l^{\prime }}+\boldsymbol{\tau _{n^{\prime
}}}}^{\alpha ,\beta }S_{ln}^{\alpha }S_{l^{\prime }n^{\prime }}^{\beta }
\label{1}
\end{eqnarray}

where $J_{\mathbf{R}_{l}+\boldsymbol{\tau }_{n},\mathbf{R}_{l^{\prime }}+%
\boldsymbol{\tau _{n^{\prime }}}}^{\alpha ,\beta }$, a $3\times 3$ tensor,
represents the spin exchange parameter. Here $\mathbf{R}_{l}$ and $%
\boldsymbol{\tau }_{n}$ represent the lattice translation vector and the
position of magnetic ions in the lattice basis, while $\alpha $ and $\beta $
denote $x,y$ or $z$ the cartesian components. As a $3\times 3$ real tensor, $%
\mathbf{J}_{\mathbf{R}_{l}+\boldsymbol{\tau }_{n},\mathbf{R}_{l^{\prime }}+%
\boldsymbol{\tau _{n^{\prime }}}}$ could be expanded as three terms, and Eq.
(\ref{1}) could be written as

\begin{eqnarray}
H &=&\sum_{l,n,l^{\prime },n^{\prime }}J_{\mathbf{R}_{l}+\boldsymbol{\tau }%
_{n},\mathbf{R}_{l^{\prime }}+\boldsymbol{\tau _{n^{\prime }}}}\mathbf{S}%
_{ln}\cdot \mathbf{S}_{l^{\prime }n^{\prime }}  \notag \\
&&+\sum_{l,n,l^{\prime },n^{\prime }}\mathbf{D}_{\mathbf{R}_{l}+\boldsymbol{%
\tau }_{n},\mathbf{R}_{l^{\prime }}+\boldsymbol{\tau _{n^{\prime }}}}\cdot
\lbrack \mathbf{S}_{ln}\times \mathbf{S}_{l^{\prime }n^{\prime }}]  \notag \\
&&+\sum_{l,n,l^{\prime },n^{\prime }}\mathbf{S}_{ln}\cdot \Gamma _{\mathbf{R}%
_{l}+\boldsymbol{\tau }_{n},\mathbf{R}_{l^{\prime }}+\boldsymbol{\tau
_{n^{\prime }}}}\cdot \mathbf{S}_{l^{\prime }n^{\prime }}  \label{1-2}
\end{eqnarray}

Here the first term describes the isotropic Heisenberg Hamiltonian with the
scalar term $J_{\mathbf{R}_{l}+\boldsymbol{\tau }_{n},\mathbf{R}_{l^{\prime
}}+\boldsymbol{\tau _{n^{\prime }}}}$, the second one represents the
antisymmetric Dzyaloshinskii-Moriya (DM) interactions with the vector term $%
\mathbf{D}_{\mathbf{R}_{l}+\boldsymbol{\tau }_{n},\mathbf{R}_{l^{\prime }}+%
\boldsymbol{\tau _{n^{\prime }}}}$ \cite{Dzyaloshinsky,dm}, and the third
one is the rest of anisotropic terms with the symmetric tensor term $\Gamma
_{\mathbf{R}_{l}+\boldsymbol{\tau }_{n},\mathbf{R}_{l^{\prime }}+\boldsymbol{%
\tau _{n^{\prime }}}}$ \cite{dm}. It is commonly believed that the magnitude
of the DM interaction and $\Gamma _{\mathbf{R}_{l}+\boldsymbol{\tau }_{n},%
\mathbf{R}_{l^{\prime }}+\boldsymbol{\tau _{n^{\prime }}}}$ are proportional
to spin-orbit coupling (SOC) strength $\lambda $ and $\lambda^{2} $,
respectively \cite{dm}. For the materials with large $\lambda$, such as
\emph{f} electronic systems, multipolar interactions may become important
\cite{multipolarnew}, thus we restrict us on the cases with small $\lambda$
and ignore the third term in Eq. (\ref{1-2}) \footnote{%
Eq. (\ref{1}) may also not be suitable for the case with orbitally degenerate
\cite{KK-1982}}. To take account for non-collinear cases, we use the polar
angle $\theta _{n}$ and azimuthal angle $\phi _{n}$ for the spin orientation
of magnetic ion at $n$ site.

Following the LSWT \cite{spinwavebook}, we perform the Holstein-Primakoff
transformation and the Fourier transformation, and the Eq. (\ref{1}) could
be written as

\begin{equation}
\sum_{k}\psi ^{\dag }(\mathbf{k})H(\mathbf{k})\psi (\mathbf{k})  \label{2}
\end{equation}

where $\psi ^{\dag }(\mathbf{k})=[a_{1}^{\dag }(\mathbf{k}),...,a_{i}^{\dag
}(\mathbf{k}),...,a_{N}^{\dag }(\mathbf{k}),a_{1}(-\mathbf{k}),...,$ $a_{i}(-%
\mathbf{k}),...,a_{N}(-\mathbf{k})]$, in which $a_{i}^{\dag }(\mathbf{k})$
and $a_{i}(\mathbf{k})$ represent the canonical boson creation and
annihilation operators with wave vector $\mathbf{k}$. Here $i$ runs from 1
to N, and N is the number magnetic ions per unit cells. The Hermitian matrix
$H(\mathbf{k})$ in Eq. (\ref{2}) is expressed as

\begin{equation}
H(\mathbf{k})=%
\begin{bmatrix}
h(\mathbf{k}) & h{^{\prime }}(\mathbf{k}) \\
h{^{\prime }}(\mathbf{k}){^{\dag }} & h(-\mathbf{k}){^{\top }}%
\end{bmatrix}
\label{ourHk}
\end{equation}

\bigskip Here $h(\mathbf{k})$ and $h{^{\prime }}(\mathbf{k})$\ are expressed
by

\begin{eqnarray}
h(\mathbf{k})_{n,n^{\prime }} &=&\sum_{l}S(A_{n,n^{\prime }}J_{\boldsymbol{%
\tau }_{n},\boldsymbol{\tau }_{n^{\prime }}+\mathbf{R}_{l}}+\mathbf{O}%
_{n,n^{\prime }}\cdot \mathbf{D}_{\boldsymbol{\tau }_{n},\boldsymbol{\tau }%
_{n^{\prime }}+\mathbf{R}_{l}})  \notag \\
&&e^{i\mathbf{k}\cdot \mathbf{R}_{l}}-\delta _{n,n^{\prime
}}\sum_{l,n^{\prime \prime }}S(B_{n,n^{\prime \prime }}J_{\boldsymbol{\tau }%
_{n},\boldsymbol{\tau }_{n^{\prime \prime }}+\mathbf{R}_{l}}  \notag \\
&&+\mathbf{P}_{n,n^{\prime \prime }}\cdot \mathbf{D}_{\boldsymbol{\tau }_{n},%
\boldsymbol{\tau }_{n^{\prime \prime }}+\mathbf{R}_{l}})  \notag \\
h{^{\prime }}(\mathbf{k})_{n,n^{\prime }} &=&\sum_{l}S(C_{n,n^{\prime }}J_{%
\boldsymbol{\tau }_{n},\boldsymbol{\tau }_{n^{\prime }}+\mathbf{R}_{l}}+%
\mathbf{Q}_{n,n^{\prime }}\cdot \mathbf{D}_{\boldsymbol{\tau }_{n},%
\boldsymbol{\tau }_{n^{\prime }}+\mathbf{R}_{l}})  \notag \\
&&e^{i\mathbf{k}\cdot \mathbf{R}_{l}}  \label{hk}
\end{eqnarray}

where $\delta _{n,n^{\prime }}$ is the Kronecker delta function, while $%
A_{n,n^{\prime }}$, $B_{n,n^{\prime }}$, $C_{n,n^{\prime }}$, $\mathbf{O}%
_{n,n^{\prime }}$, $\mathbf{P}_{n,n^{\prime }}$ and $\mathbf{Q}_{n,n^{\prime
}}$ are parameters related to the spin directions at $n$ and $n^{\prime }$
sites (see SM for details).

Considering the commutation relation of $\psi (\mathbf{k})$ and $\psi ^{\dag
}(\mathbf{k})$, we need perform the following transformation (see SM for
details):

\begin{equation}
H_{J}(\mathbf{k})=I_{-}H(\mathbf{k})  \label{3}
\end{equation}

Through numerically diagonalizing the $H_{J}(\mathbf{k})$ in Eq. (\ref{3}),
we can obtain the magnon energies $\omega _{i}(\mathbf{k}) (i=1,...,N)$ at
wave vector $k$. In contrary without diagonalization, SSME can be
analytically expressed as:

\begin{eqnarray}
\sum_{i}\omega _{i}^{2}(\mathbf{k}) &=&\frac{1}{2}Tr([H_{J}(\mathbf{k})]^{2})
\notag \\
&=&\frac{1}{2}Tr[h^{2}(\mathbf{k})+h^{2}(-\mathbf{k}){^{\top }}-h{^{\prime }}%
(\mathbf{k})h{^{\prime }}(\mathbf{k}){^{\dag }}  \notag \\
&&{-h{^{\prime }}(\mathbf{k}){^{\dag }}h{^{\prime }}(\mathbf{k})}]
\label{SSME}
\end{eqnarray}

As shown in Eq. (\ref{hk}), $h(\mathbf{k})$ and $h{^{\prime }}(\mathbf{k})$
basically depend on the orientation of magnetic moments and the Fourier
transformation of MEIs $\mathbf{J}_{\mathbf{R}_{l}+\boldsymbol{\tau }_{n},%
\mathbf{\ R}_{l^{\prime }}+\boldsymbol{\tau _{n^{\prime }}}}$ and $\mathbf{D}%
_{\mathbf{R}_{l}+\boldsymbol{\tau }_{n},\mathbf{R}_{l^{\prime }}+\boldsymbol{%
\tau _{n^{\prime }}}}$. Thus, for arbitrary $k$, $\sum_{i}\omega _{i}^{2}(%
\mathbf{k})$ can be expressed by a quadratic polynomial of MEIs. With the
assumption of $R_{cut}$, which related with how many MEIs had been
considered 
, one can obtain simple relationships between SSME at different wave vectors
\emph{k}.

\section{Results and discussion}

\bigskip
\begin{table}[tbph]
\caption{The WPs and the coordinates of the 8 magnetic ions in the
conventional unit cell basis vectors\ for the example shown here. The
Wyckoff positions are labeled in the space group P4/n (SG 85). The polar
angles $\protect\theta _{n}$\ and azimuthal angles $\protect\phi _{n}$\ of
these magnetic ions in our selected collinear and non-collinear states are
also provided. }
\label{positions}\centering%
\begin{tabular}{ccccc}
\hline\hline
WP & $n$ & $\tau _{n}$ & \multicolumn{2}{c}{$(\theta _{n},\phi _{n})$} \\
&  &  & collinear & non-collinear \\ \hline
4d & 1 & (0, 0, 0) & (0,0) & $(\theta ,\pi /2)$ \\
& 2 & (0.5, 0, 0) & (0,0) & $(\theta ,-\pi /2)$ \\
& 3 & (0, 0.5, 0) & (0,0) & $(\theta ,\pi /2)$ \\
& 4 & (0.5, 0.5, 0) & (0,0) & $(\theta ,-\pi /2)$ \\ \hline
2a & 5 & (0.75, 0.25, 0) & (0,0) & (0,0) \\
& 6 & (0.25, 0.75, 0) & (0,0) & (0,0) \\ \hline
2c & 7 & (0.25, 0.25, 0.1) & (0,0) & (0,0) \\
& 8 & (0.75, 0.75, $-$0.1) & (0,0) & (0,0) \\ \hline
\end{tabular}%
\end{table}

\begin{table}[tbph]
\caption{The bonds for the 1st, 2nd and 3rd NNs of the example shown in here
and the corresponding Heisenberg exchange interactions of the collinear FM
configuration (i.e. the case with BNS 85.59 ). Each bond is characterized by
the positions of the two endings: $\boldsymbol{\protect\tau }_{n},%
\boldsymbol{\protect\tau }_{n^{\prime }}+\mathbf{R}_{l}$. The unit of
distance is taken as the lattice constant $a$.}
\label{1NN}\centering%
\begin{tabular}{ccccc}
\hline\hline
& distance(a) & $n$ & $n^{\prime }$ & $R_{l}$ \\ \hline
$J_{1}$ & 0.35 & 1 & 5 & $(-1,0,0)$ \\
&  & 1 & 6 & $(0,-1,0)$ \\
&  & 2 & 5 & $(0,0,0)$ \\
&  & 2 & 6 & $(0,-1,0)$ \\
&  & 3 & 5 & $(-1,0,0)$ \\
&  & 3 & 6 & $(0,0,0)$ \\
&  & 4 & 5 & $(0,0,0)$ \\
&  & 4 & 6 & $(0,0,0)$ \\ \hline
$J_{2}$ & 0.36 & 1 & 7 & $(0,0,0)$ \\
&  & 1 & 8 & $(-1,-1,0)$ \\
&  & 2 & 7 & $(0,0,0)$ \\
&  & 2 & 8 & $(0,-1,0)$ \\
&  & 3 & 7 & $(0,0,0)$ \\
&  & 3 & 8 & $(-1,0,0)$ \\
&  & 4 & 7 & $(0,0,0)$ \\
&  & 4 & 8 & $(0,0,0)$ \\ \hline
$J_{3}$ & 0.5 & 1 & 2 & $(0,0,0)$ \\
&  & 1 & 2 & $(-1,0,0)$ \\
&  & 1 & 3 & $(0,0,0)$ \\
&  & 1 & 3 & $(0,-1,0)$ \\
&  & 2 & 4 & $(0,0,0)$ \\
&  & 2 & 4 & $(0,-1,0)$ \\
&  & 3 & 4 & $(0,0,0)$ \\
&  & 3 & 4 & $(-1,0,0)$ \\ \hline\hline
\end{tabular}%
\end{table}

\begin{table*}[tbph]
\caption{The obtained SSME relationship of the collinear FM example shown in
here (i.e. the case with BNS 85.59). The first column $J_{x}$ represents up
to the $x$-th NN MEI. The coordinate of six high-symmetry $k$ points: $%
\Gamma (0,0,0)$, $X(\frac{1}{2},0,0)$, $M(\frac{1}{2},\frac{1}{2},0)$, $%
Z(0,0,\frac{1}{2})$, $R(\frac{1}{2},0,\frac{1}{2})$, and $A(\frac{1}{2},%
\frac{1}{2},\frac{1}{2})$.}
\label{relation1}\centering%
\begin{tabular}{ll}
\hline\hline
$J_{\max }$ & relation \\ \hline
$J_{2}$ & $\sum_{i}\omega _{i}^{2}(k)=C$ \\ \hline
$J_{4}$ & $%
\begin{array}{l}
\sum_{i}\omega _{i}^{2}(\Gamma )=\sum_{i}\omega _{i}^{2}(Z) \\
\sum_{i}\omega _{i}^{2}(X)=\sum_{i}\omega _{i}^{2}(R) \\
\sum_{i}\omega _{i}^{2}(M)=\sum_{i}\omega _{i}^{2}(A) \\
\multicolumn{1}{c}{2\sum_{i}\omega _{i}^{2}(X)=\sum_{i}\omega
_{i}^{2}(\Gamma )+\sum_{i}\omega _{i}^{2}(M)}%
\end{array}%
$ \\ \hline
$J_{12}$ & $%
\begin{array}{l}
\sum_{i}\omega _{i}^{2}(\Gamma )=\sum_{i}\omega _{i}^{2}(Z) \\
\sum_{i}\omega _{i}^{2}(X)=\sum_{i}\omega _{i}^{2}(R) \\
\sum_{i}\omega _{i}^{2}(M)=\sum_{i}\omega _{i}^{2}(A)%
\end{array}%
$ \\ \hline
$J_{15}$ & $\sum_{i}\omega _{i}^{2}(\Gamma )-\sum_{i}\omega
_{i}^{2}(Z)=\sum_{i}\omega _{i}^{2}(X)-\sum_{i}\omega
_{i}^{2}(R)=\sum_{i}\omega _{i}^{2}(M)-\sum_{i}\omega _{i}^{2}(A){}$ \\
\hline
$J_{22}$ & $2\sum_{i}\omega _{i}^{2}(X)-\sum_{i}\omega _{i}^{2}(\Gamma
)-\sum_{i}\omega _{i}^{2}(M)=2\sum_{i}\omega _{i}^{2}(R)-\sum_{i}\omega
_{i}^{2}(Z)-\sum_{i}\omega _{i}^{2}(A)$ \\ \hline
\end{tabular}%
\end{table*}

We illustrate the usage of our results by following typical example. Without
loss of generality, we choose space group P4/n (SG 85) to present our
discussion and set the ratio between lattice constant $c/a$ as 0.8. We put
the magnetic ions at three nonequivalent crystallographic sites: $4d$ (0, 0,
0), $2a$ (0.25, 0.75, 0) and $2c$ (0.25, 0.25, $z$) Wyckoff positions (WPs),
as summarized in Table \ref{positions}. While the $4d$ and $2a$ WPs had been
completely determined by the spatial symmetry, the coordinates of $2c$ WP
have a variable $z$ and here we adopt it as $z=0.1$. There are two
generators for this space group: the four-fold rotation $%
\{4_{001}^{+}|1/2,0,0\}$ and inversion operation $\{\overline{1}|0,0,0\}$,
where the left part represents the rotation, the right part means the
lattice translation, and $\overline{1}$ denotes the inversion symmetry. We
firstly consider the most simple case: isotropic Heisenberg model with all
the spins along $z$. Considering the orientations of the magnetic moments,
the space group could be divided into four types of magnetic space groups
\footnote{%
Type-I magnetic space group has no any additional symmetry compared with the
corresponding space group, while type-II magnetic space group has an
additional anti-symmetry version of every symmetry operation. For type-III
magnetic space group, there are additional anti-symmetry versions for the
half of its symmetry operations. Specially, type-IV magnetic space group has
additional combined spatial translation-time reversal symmetry.}. The case
with this collinear ferromagnetic (FM) ordering belongs to the type-I
magnetic space group (BNS 85.59), and its magnetic configuration does not
reduce the spatial symmetry. Since all the spins along \emph{z} direction,
polar angle $\theta _{n}$ and azimuthal angle $\phi _{n}$ are equal to 0 as
listed in Table \ref{positions}, thus according to the Eq. (\ref{fma}-\ref%
{fmc}) in SM, the parameters $A_{n,n^{\prime }}$, $B_{n,n^{\prime }}$ and $%
C_{n,n^{\prime }}$ in Eq. (\ref{hk}) for this collinear FM state\ becomes 1,
1 and 0 respectively. Consequently, the SSME at wave vector $k$ could be
written as

\begin{widetext}
\begin{eqnarray}
\sum_{i}\omega _{i}^{2}(\mathbf{k}) &=&\frac{1}{2}Tr([H_{J}(\mathbf{k})]^{2})
\notag \\
&=&S^{2}\sum_{n\neq n^{\prime },l,l^{\prime }}J_{\boldsymbol{\tau }_{n},%
\boldsymbol{\tau }_{n^{\prime }}+\mathbf{R}_{l^{\prime }}}J_{\boldsymbol{%
\tau }_{n},\boldsymbol{\tau }_{n^{\prime }}+\mathbf{R}_{l}}e^{i\mathbf{k}%
\cdot (\mathbf{R}_{l}-\mathbf{R}_{l^{\prime }})}+S^{2}\sum_{n}\left[
\sum_{n^{\prime \prime },l}J_{\boldsymbol{\tau }_{n},\boldsymbol{\tau }%
_{n^{\prime \prime }}+\mathbf{R}_{l}}+\sum_{l}J_{\boldsymbol{\tau }_{n},%
\boldsymbol{\tau }_{n}+\mathbf{R}_{l}}e^{i\mathbf{k}\cdot \mathbf{R}_{l}}%
\right] ^{2}  \label{fm001}
\end{eqnarray}%
\end{widetext}

As shown in the above formula, the key for SSME is the exchange path between
magnetic ions $\boldsymbol{\tau }_{n}$ and $\boldsymbol{\tau }_{n^{\prime }}+%
\mathbf{R}_{l^{\prime }}$, and the related MEI $J_{\boldsymbol{\tau }_{n},%
\boldsymbol{\tau }_{n^{\prime }}+\mathbf{R}_{l^{\prime }}}$. As shown in
Table \ref{1NN}, the first and second nearest neighbor have similar distances
(0.35 $a$ vs 0.36 $a$, $a$ is lattice parameter). Crystal symmetry imposes
strong restrictions on the MEIs as shown in SM, and according to the spatial
symmetry in this space group, all the first nearest neighbor exchange paths
have the same MEI value, and we denotes it as $J_{1}$, same as that, we can
label all the second nearest neighbor MEI as $J_{2}$. Considering only the
first two NN interactions $J_{1}$ and $J_{2}$, the term of $J_{\boldsymbol{%
\tau }_{n},\boldsymbol{\tau }_{n}+\mathbf{R}_{l}}$ does not exist as shown
in Table \ref{1NN}, and the only $k$ dependence of SSME comes from the first
term in Eq. (\ref{fm001}). Namely, we need to check the non-zero MEIs $J_{%
\boldsymbol{\tau }_{n},\boldsymbol{\tau }_{n^{\prime }}+\mathbf{R}%
_{l^{\prime }}}$ and $J_{\boldsymbol{\tau }_{n},\boldsymbol{\tau }%
_{n^{\prime }}+\mathbf{R}_{l}}$, with the requirement of $\mathbf{R}_{l}\neq
\mathbf{R}_{l^{\prime }}$. As clearly shown in Table \ref{1NN}, such kind of
exchange path is also not exist. Thus, although spin wave has dispersion at
the entire BZ, we get a surprisingly simple result of $\sum_{i}\omega
_{i}^{2}(\mathbf{k})=C$ with considering only $J_{1}$ and $J_{2}$.

We further take into account the impact of longer-ranged exchange paths.
With the third NN MEI $J_{3}$ been considered, there exist more than one
exchange paths connect a pair of $\boldsymbol{\tau }_{n}$ and $\boldsymbol{%
\tau }_{n^{\prime }}+\mathbf{R}_{l}$. For example, both ${\boldsymbol{\tau }%
_{1},\boldsymbol{\tau }_{2}}$ pair and ${\boldsymbol{\tau }_{1},\boldsymbol{%
\tau }_{2}+\mathbf{R}_{-100}}$ pair belong to $J_{3}$ exchange path as shown
in Table \ref{1NN}. As a result, $\sum_{i}\omega _{i}^{2}(\mathbf{k})$ is no
longer equal to constant. Thus, if the observed SSME shows very weak $k$
dependence, one can asserts that the MEIs beyond $J_{2}$ are ignorable.
Since for the high symmetry $k$ points at BZ, $e^{i\mathbf{k}\cdot (\mathbf{R%
}_{l}-\mathbf{R}_{l^{\prime }})}$ usually has simple values (equals to $\pm
1 $ in this magnetic system), one can expect simple relation between SSME at
these $k$ points. We indeed get several simple relationships with the MEIs
up to $J_{3}$ : $\sum_{i}\omega _{_{i}}^{2}(\Gamma ){}=\sum_{i}\omega
_{_{i}}^{2}(Z)$, $\sum_{i}\omega _{_{i}}^{2}(X)=\sum_{i}\omega
_{_{i}}^{2}(R) $, $\sum_{i}\omega _{_{i}}^{2}(M){}=\sum_{i}\omega
_{_{i}}^{2}(A)$ and $2\sum_{i}\omega _{_{i}}^{2}{(X)}=\sum_{i}\omega
_{_{i}}^{2}(\Gamma )+\sum_{i}\omega _{_{i}}^{2}(M)$. It is interest to see
that these four simple relations about SSME remains after including the
fourth nearest neighbor MEI $J_{4}$. The algorithm about SSME is simple,
which allow us quickly analyze the effect of considering further MEI. We
summarize the results at Table \ref{relation1}. Using the measured magnon
energies at only six high-symmetry $k$ points, one can unambiguous determine
the real-space range within which the MEI with considerable value based on
the Table \ref{relation1}.

After collinear FM configuration, we now illustrate the applications for
non-collinear case. We still use the crystal structure mentioned above and
fix the magnetic moments at 2\emph{a} and 2\emph{c} WPs still along \emph{z}
direction. While the azimuthal angle $\phi $ for magnetic moments at 4\emph{d%
} WP is $\pm \pi $, we assume their polar angles as a free parameter $\theta
$ as shown in the Table \ref{positions}. Although an isotropic Heisenberg
Hamiltonian may not produce the above non-collinearity, we still use it to
demonstrate our method and show the discussion about anisotropic spin model
at later. This non-collinear magnetic state belongs to the Type-I magnetic
space group BNS 13.65. While the inversion symmetry $\{\overline{1}|0,0,0\}$
are maintained, the the deviation from $z$ direction reduces the four-fold
rotation symmetry $\{4_{001}^{+}|1/2,0,0\}$ to the two-fold rotation
operation $\{2_{001}|1/2,1/2,0\}$. As the result, many symmetry-related
exchange pathes in collinear spin ordering case become inequivalent. For
example, as shown in Table \ref{non1} of SM, the eight first NN exchange
paths in collinear spin ordering are no longer equivalent and divided into
two groups, which are labeled as $J_{1}$ and$\ J_{2}$ for this
non-collinear magnetic case. The parameters in Eq. (\ref{hk}) are also
depended on the magnetic moment directions, thus non-collinearity results in
different relationship between SSME, which are listed in Table \ref{non2new}%
. 
We also want to mention that for the localized magnetic systems, the MEIs
should not be sensitive to the magnetic configurations as also required by
energy-mapping method for calculating MEIs. For such cases, one can still
use the symmetry operations in collinear instead of in non-collinear case to determine equivalent
exchange path. Namely if this non-collinear magnetism is very localized, the
MEIs will still approximately satisfy the Table \ref{1NN}. Based on Table %
\ref{1NN} (i.e. the right part of the Table \ref{non1} in SM) instead of the
left part of the Table \ref{non1} in SM, we applying our method to this
non-collinear case with localized magnetism and list the results in the
right part of Table \ref{non2new}. As shown in Table \ref{non2new}, the free
parameter $\theta $ about the magnetic moments directions explicitly appear
in the relationship about SSME. Thus, for localized non-collinear magnetic
materials, one may determine the magnetic moments directions based on the
magnon energies at three wave vectors (i.e. $\Gamma$, $X$ and $M$) in the
case that MEIs further than $J_{3}$ are ignorable.

It is worth to mention that our method is also valid for the materials with
considerable DM interactions \cite{Dzyaloshinsky,dm}. One can still calculate
SSME by Eq. (\ref{SSME}) and directly use the program provided in SM to
explore the relationship between them. The magnetic anisotropy may also
comes from the SIA \cite{book-1,book-2,book-3}. With the SIA considered, the
Hamiltonian becomes $H_{total}=H+H_{SIA}$, here $H$ is the term shown in Eq.
(\ref{1-2}) while $H_{SIA}$ represents the term of SIA. Here we adopt a
popular form $H_{SIA}=\sum_{l,n}K(S_{ln}^{z})^{2}$ \ \cite{book-3} where $K$
is the strength of SIA. Based on the standard LSWT, one can easily obtain
the spin Hamiltonian at arbitrary wave vector $k$ to be $%
H_{total}(k)=H_{J}(k)+2SKI_{-}$. Adding SIA term into the case of Heisenberg
model with collinear FM magnetic ordering given in this work, the SSME could
be written as

\begin{eqnarray}
\sum_{i}\omega _{i}^{2}(\mathbf{k}) &=&\frac{1}{2}Tr[(H_{total}(\mathbf{k}%
))^{2}]  \notag \\
&=&\frac{1}{2}Tr[(H_{J}(\mathbf{k}))^{2}]+4NS^{2}K^{2}+4S^{2}K\cdot  \notag
\\
&&\sum_{n}\left[ \sum_{n^{\prime \prime },l}J_{\boldsymbol{\tau }_{n},%
\boldsymbol{\tau }_{n^{\prime \prime }}+\mathbf{R}_{l}}+\sum_{l}J_{%
\boldsymbol{\tau }_{n},\boldsymbol{\tau }_{n}+\mathbf{R}_{l}}e^{i\mathbf{k}%
\cdot \mathbf{R}_{l}}\right]  \notag \\
&&  \label{SIA}
\end{eqnarray}

Based on Eq. (\ref{SIA}), one can easily prove that including SIA will not
affect the results given in Table \ref{relation1}. For the other cases, one
can simply use our code which has implemented effect of SIA to obtain the
results.

\begin{table*}[tbph]
\caption{The obtained results about SSME for the non-collinear case shown in
this work (i.e. the case with symmetry of BNS 13.65). We also give the
result for the same non-collinear configuration with the magnetism is very
localized in the right part. For the localized magnetism case, the MEIs are
not sensitive to the spin ordering, thus one can use the symmetry of SG 85
to determine if the exchange pathes have the same MEIs, namely use the
results given in the right part of Table \protect\ref{non1} in SM. The $x$ in $%
J_{x}$ still represents up to the $x$-th NN MEI, and the coordination of six
high-symmetry points had been shown in Table \protect\ref{relation1}.}
\label{non2new}\centering%
\begin{tabular}{ll|ll}
\hline\hline
$J_{\max }$ & relation & $J_{\max }$ & relation \\ \hline
$J_{4}$ & $\sum_{i}\omega _{i}^{2}(k)=C$ & $J_{2}$ & $\sum_{i}\omega
_{i}^{2}(k)=C$ \\ \hline
$J_{5}$ & $%
\begin{array}{l}
\sum_{i}\omega _{i}^{2}(\Gamma )=\sum_{i}\omega _{i}^{2}(Z) \\
\multicolumn{1}{c}{\sum_{i}\omega _{i}^{2}(X)=\sum_{i}\omega
_{i}^{2}(R)=\sum_{i}\omega _{i}^{2}(M)=\sum_{i}\omega _{i}^{2}(A)}%
\end{array}%
$ & $J_{3}$ & $%
\begin{array}{l}
\sum_{i}\omega _{i}^{2}(\Gamma )=\sum_{i}\omega _{i}^{2}(Z) \\
\sum_{i}\omega _{i}^{2}(X)=\sum_{i}\omega _{i}^{2}(R) \\
\sum_{i}\omega _{i}^{2}(M)=\sum_{i}\omega _{i}^{2}(A) \\
\multicolumn{1}{c}{(1+\cos 2\theta )\sum_{i}\omega _{i}^{2}(X)=\omega
_{i}^{2}(\Gamma )+\cos 2\theta \sum_{i}\omega _{i}^{2}(M)}%
\end{array}%
$ \\ \hline
$J_{24}$ & $%
\begin{array}{l}
\sum_{i}\omega _{i}^{2}(\Gamma )=\sum_{i}\omega _{i}^{2}(Z) \\
\sum_{i}\omega _{i}^{2}(X)=\sum_{i}\omega _{i}^{2}(R) \\
\sum_{i}\omega _{i}^{2}(M)=\sum_{i}\omega _{i}^{2}(A)%
\end{array}%
$ & $J_{12}$ & $%
\begin{array}{l}
\sum_{i}\omega _{i}^{2}(\Gamma )=\sum_{i}\omega _{i}^{2}(Z) \\
\sum_{i}\omega _{i}^{2}(X)=\sum_{i}\omega _{i}^{2}(R) \\
\sum_{i}\omega _{i}^{2}(M)=\sum_{i}\omega _{i}^{2}(A)%
\end{array}%
$ \\ \hline
$J_{28}$ & $%
\begin{array}{c}
\sum_{i}\omega _{i}^{2}(\Gamma )-\sum_{i}\omega _{i}^{2}(Z)=\sum_{i}\omega
_{i}^{2}(X)-\sum_{i}\omega _{i}^{2}(R) \\
=\sum_{i}\omega _{i}^{2}(M){}-\sum_{i}\omega _{i}^{2}(A)%
\end{array}%
$ & $J_{15}$ & $%
\begin{array}{c}
\sum_{i}\omega _{i}^{2}(\Gamma )-\sum_{i}\omega _{i}^{2}(Z)=\sum_{i}\omega
_{i}^{2}(X)-\sum_{i}\omega _{i}^{2}(R) \\
=\sum_{i}\omega _{i}^{2}(M){}-\sum_{i}\omega _{i}^{2}(A)%
\end{array}%
$ \\ \hline
&  & $J_{19}$ & $%
\begin{array}{c}
2\sum_{i}\omega _{i}^{2}(X)-\sum_{i}\omega _{i}^{2}(\Gamma )-\sum_{i}\omega
_{i}^{2}(M) \\
=2\sum_{i}\omega _{i}^{2}(R)-\sum_{i}\omega _{i}^{2}(Z)-\sum_{i}\omega
_{i}^{2}(A)%
\end{array}%
$ \\ \hline
\end{tabular}%
\end{table*}

\section{Conclusion}

In summary, appropriate magnetic model play crucial role in investigating
various magnetic properties. Unfortunately the current methods for
extracting MEIs face a severe limitation about how many MEIs need to be
included in the spin Hamiltonian. In this work, we circumvent this
methodological bottleneck by noticing that for quadratic spin Hamiltonian,
there is a simple connection between SSME and the considered MEIs. Namely,
there is $R_{cut}$-related rules between SSME at high-symmetry points.
By efficient measurements of magnon energies only at several high-symmetry $k$ points,
one can check up to
which $R_{cut}$, the experimental SSME start to deviate from the obtained $%
R_{cut}$-related rules, and subsequently determine the real-space
range beyond which MEIs can be safely neglected. For the
localized non-collinear magnetic systems, our results may also be used to
determine the directions of magnetization. We also provide a program, directly
utilizing it, one can get the relationship of SSME for any crystal magnetic
materials described by Hamiltonian with pairwise spin. Besides the well used
symmetry analysis for the symmetry-related $k$ points, we expect that
similar generic $R_{cut}$-sensitive rules also exist in other Hamiltonian
with only quadratic Fermi or boson operators.

\section{Acknowledgements}

This work was supported by the NSFC (No. 11834006, 12004170, 51721001, and
11790311), National Key R\&D Program of China (No. 2018YFA0305704 and
2017YFA0303203) and the excellent programme in Nanjing University. Xiangang
Wan also acknowledges the support from the Tencent Foundation through the
XPLORER PRIZE.

\section{Supplemental Materials}

\subsection{The symmetry restrictions on the magnetic interactions}

As mentioned in the maintext, the microscopic magnetic model with proper
parameters is extremely important. Note that the crystal symmetry impose
restrictions on the magnetic model and its parameters. Here we consider a
general pairwise spin model as shown in the maintext

\begin{equation}
H=\sum_{l,n,l^{\prime },n^{\prime }}\mathbf{S}_{ln}\cdot \mathbf{J}_{\mathbf{%
R}_{l}+\boldsymbol{\tau }_{n},\mathbf{R}_{l^{\prime }}+\boldsymbol{\tau
_{n^{\prime }}}}\cdot \mathbf{S}_{l^{\prime }n^{\prime }}
\end{equation}

where $\mathbf{J}_{\mathbf{R}_{l}+\boldsymbol{\tau }_{n},\mathbf{R}%
_{l^{\prime }}+\boldsymbol{\tau _{n^{\prime }}}}$, a $3 \times 3$ tensor,
represents the spin exchange parameters. $R_{l}$ and $\tau _{n}$
represent the lattice translation vector and the position of magnetic ions in
the lattice basis, and $\mathbf{S}_{l^{\prime }n^{\prime }}$ means the spin
at the site of $\mathbf{R}_{l}+\boldsymbol{\tau }_{n}$ Translation symmetry
will restrict $\mathbf{J}_{\mathbf{R}_{l}+\boldsymbol{\tau }_{n},\mathbf{R}%
_{l^{\prime }}+\boldsymbol{\tau _{n^{\prime }}}}$ to be only related to $%
\mathbf{J}_{\boldsymbol{\tau }_{n},\boldsymbol{\tau _{n^{\prime }}+}\mathbf{R%
}_{l^{\prime \prime }}}$ where $R_{l^{\prime \prime }}=R_{l^{\prime }}-R_{l}$%
, irrespective of the starting unit cell. Other spatial symmetries will also
give restrictions on the magnetic exchange interactions (MEIs). We consider
a general space group element $\{\alpha |\mathbf{t}\}$, where the left part
represents the rotation and the right part means the lattice translation.
Supposing under this symmetry operator, $\mathbf{R}_{m}+\boldsymbol{\tau }%
_{p}$ and $\mathbf{R}_{m^{\prime }}+\boldsymbol{\tau _{p^{\prime }}}$
transfer to $\mathbf{R}_{l}+\boldsymbol{\tau }_{n}$ and $\mathbf{R}%
_{l^{\prime }}+\boldsymbol{\tau _{n^{\prime }}}$, respectively, meanwhile
the transformation of spin becomes $\mathbf{S}_{mp}=M(\alpha )\mathbf{S}%
_{ln} $, where $M(\alpha )$ is the representation matrix of the proper
rotation part of the operation $\alpha $ in the coordinate system, we get
the following expression:

\begin{eqnarray}
H &=&\sum_{l,n,l^{\prime },n^{\prime }}\mathbf{S}_{ln}\cdot \mathbf{J}_{%
\mathbf{R}_{l}+\boldsymbol{\tau }_{n},\mathbf{R}_{l^{\prime }}+\boldsymbol{%
\tau _{n^{\prime }}}}\cdot \mathbf{S}_{l^{\prime }n^{\prime }}  \notag \\
&=&\sum_{l,n,l^{\prime },n^{\prime }}\mathbf{S}_{ln}M^{\dag }(\alpha
)M(\alpha )\mathbf{J}_{\mathbf{R}_{l}+\boldsymbol{\tau }_{n},\mathbf{R}%
_{l^{\prime }}+\boldsymbol{\tau _{n^{\prime }}}}M^{\dag }(\alpha )M(\alpha )%
\mathbf{S}_{l^{\prime }n^{\prime }}  \notag \\
&=&\sum_{m,p,m^{\prime },p^{\prime }}\mathbf{S}_{mp}\cdot M(\alpha )\mathbf{J%
}_{\mathbf{R}_{l}+\boldsymbol{\tau }_{n},\mathbf{R}_{l^{\prime }}+%
\boldsymbol{\tau _{n^{\prime }}}}M^{\dag }(\alpha )\cdot \mathbf{S}%
_{m^{\prime }p^{\prime }}
\end{eqnarray}

Then the exchange interactions should satisfy the following condition:

\begin{equation}
\mathbf{J}_{\mathbf{R}_{m}+\boldsymbol{\tau }_{p},\mathbf{R}_{m^{\prime }}+%
\boldsymbol{\tau _{p^{\prime }}}}=M(\alpha )\mathbf{J}_{\mathbf{R}_{l}+%
\boldsymbol{\tau }_{n},\mathbf{R}_{l^{\prime }}+\boldsymbol{\tau _{n^{\prime
}}}}M^{\dag }(\alpha )  \label{Jrelation}
\end{equation}

After decomposing the $3 \times 3$ tensor $\mathbf{J}$ into scalar
Heisenberg term J and vector DM term D as in the maintext, we obtain the
following results:

\begin{eqnarray}
J_{\mathbf{R}_{m}+\boldsymbol{\tau }_{p},\mathbf{R}_{m^{\prime }}+%
\boldsymbol{\tau _{p^{\prime }}}} &=&J_{\mathbf{R}_{l}+\boldsymbol{\tau }%
_{n},\mathbf{R}_{l^{\prime }}+\boldsymbol{\tau _{n^{\prime }}}}  \notag \\
\mathbf{D}_{\mathbf{R}_{m}+\boldsymbol{\tau }_{p},\mathbf{R}_{m^{\prime }}+%
\boldsymbol{\tau _{p^{\prime }}}} &=&M(\alpha )\mathbf{D}_{\mathbf{R}_{l}+%
\boldsymbol{\tau }_{n},\mathbf{R}_{l^{\prime }}+\boldsymbol{\tau _{n^{\prime
}}}}  \label{Jrelation-2}
\end{eqnarray}

Meanwhile, it is should be noted that the Heisenberg and DM interactions
obey the following commutation relations

\begin{eqnarray}
J_{\mathbf{R}_{l^{\prime }}+\boldsymbol{\tau _{n^{\prime }},}\mathbf{R}_{l}+%
\boldsymbol{\tau }_{n}} &=&J_{\mathbf{R}_{l}+\boldsymbol{\tau }_{n},\mathbf{R%
}_{l^{\prime }}+\boldsymbol{\tau _{n^{\prime }}}}  \notag \\
\mathbf{D}_{\mathbf{R}_{l^{\prime }}+\boldsymbol{\tau _{n^{\prime }},}%
\mathbf{R}_{l}+\boldsymbol{\tau }_{n}} &=&-\mathbf{D}_{\mathbf{R}_{l}+%
\boldsymbol{\tau }_{n},\mathbf{R}_{l^{\prime }}+\boldsymbol{\tau _{n^{\prime
}}}}  \label{Jrelation-3}
\end{eqnarray}

\begin{table*}[tbph]
\caption{The distances and the bond information of corresponding MEIs (with $%
R_{cut}=$ 0.5 $a$) for the non-collinear magnetic example in the maintext
(i.e. the case with symmetry of BNS 13.65). We also list the results by
using the symmetry of BNS 85.59, which is applicable for the collinear FM
case as well as the non-collinear case with localized magnetism in the right
part.}
\label{non1}\centering%
\begin{tabular}{cccc|c|c|c|c}
\hline\hline
distance(a) & $n$ & $n^{\prime }$ & $R_{l}$ & \multicolumn{2}{|c|}{BNS 13.65}
& \multicolumn{2}{|c}{BNS 85.59} \\ \hline
0.35 & 1 & 5 & $(-1,0,0)$ & $J_{1}$ & $(D_{1}^{x},D_{1}^{y},D_{1}^{z})$ & $%
J_{1}$ & $(D_{1}^{x},D_{1}^{y},D_{1}^{z})$ \\
& 1 & 6 & $(0,-1,0)$ & $J_{1}$ & $(D_{1}^{x},D_{1}^{y},D_{1}^{z})$ & $J_{1}$
& $(D_{1}^{x},D_{1}^{y},D_{1}^{z})$ \\
& 2 & 5 & $(0,0,0)$ & $J_{2}$ & $(D_{2}^{x},D_{2}^{y},D_{2}^{z})$ & $J_{1}$
& $(-D_{1}^{y},D_{1}^{x},D_{1}^{z})$ \\
& 2 & 6 & $(0,-1,0)$ & $J_{2}$ & $(D_{2}^{x},D_{2}^{y},D_{2}^{z})$ & $J_{1}$
& $(-D_{1}^{y},D_{1}^{x},D_{1}^{z})$ \\
& 3 & 5 & $(-1,0,0)$ & $J_{2}$ & $(-D_{2}^{x},-D_{2}^{y},D_{2}^{z})$ & $%
J_{1} $ & $(D_{1}^{y},-D_{1}^{x},D_{1}^{z})$ \\
& 3 & 6 & $(0,0,0)$ & $J_{2}$ & $(-D_{2}^{x},-D_{2}^{y},D_{2}^{z})$ & $J_{1}$
& $(D_{1}^{y},-D_{1}^{x},D_{1}^{z})$ \\
& 4 & 5 & $(0,0,0)$ & $J_{1}$ & $(-D_{1}^{x},-D_{1}^{y},D_{1}^{z})$ & $J_{1}$
& $(-D_{1}^{x},-D_{1}^{y},D_{1}^{z})$ \\
& 4 & 6 & $(0,0,0)$ & $J_{1}$ & $(-D_{1}^{x},-D_{1}^{y},D_{1}^{z})$ & $J_{1}$
& $(-D_{1}^{x},-D_{1}^{y},D_{1}^{z})$ \\ \hline
0.36 & 1 & 7 & $(0,0,0)$ & $J_{3}$ & $(D_{3}^{x},D_{3}^{y},D_{3}^{z})$ & $%
J_{2}$ & $(D_{2}^{x},D_{2}^{y},D_{2}^{z})$ \\
& 1 & 8 & $(-1,-1,0)$ & $J_{3}$ & $(D_{3}^{x},D_{3}^{y},D_{3}^{z})$ & $J_{2}$
& $(D_{2}^{x},D_{2}^{y},D_{2}^{z})$ \\
& 2 & 7 & $(0,0,0)$ & $J_{4}$ & $(D_{4}^{x},D_{4}^{y},D_{4}^{z})$ & $J_{2}$
& $(-D_{2}^{y},D_{2}^{x},D_{2}^{z})$ \\
& 2 & 8 & $(0,-1,0)$ & $J_{4}$ & $(D_{4}^{x},D_{4}^{y},D_{4}^{z})$ & $J_{2}$
& $(-D_{2}^{y},D_{2}^{x},D_{2}^{z})$ \\
& 3 & 7 & $(0,0,0)$ & $J_{4}$ & $(-D_{4}^{x},-D_{4}^{y},D_{4}^{z})$ & $J_{2}$
& $(D_{2}^{y},-D_{2}^{x},D_{2}^{z})$ \\
& 3 & 8 & $(-1,0,0)$ & $J_{4}$ & $(-D_{4}^{x},-D_{4}^{y},D_{4}^{z})$ & $%
J_{2} $ & $(D_{2}^{y},-D_{2}^{x},D_{2}^{z})$ \\
& 4 & 7 & $(0,0,0)$ & $J_{3}$ & $(-D_{3}^{x},-D_{3}^{y},D_{3}^{z})$ & $J_{2}$
& $(-D_{2}^{x},-D_{2}^{y},D_{2}^{z})$ \\
& 4 & 8 & $(0,0,0)$ & $J_{3}$ & $(-D_{3}^{x},-D_{3}^{y},D_{3}^{z})$ & $J_{2}$
& $(-D_{2}^{x},-D_{2}^{y},D_{2}^{z})$ \\ \hline
0.5 & 1 & 2 & $(0,0,0)$ & $J_{5}$ & $(D_{5}^{x},D_{5}^{y},D_{5}^{z})$ & $%
J_{3}$ & $(D_{3}^{x},D_{3}^{y},D_{3}^{z})$ \\
& 1 & 2 & $(-1,0,0)$ & $J_{5}$ & $(D_{5}^{x},D_{5}^{y},D_{5}^{z})$ & $J_{3}$
& $(D_{3}^{x},D_{3}^{y},D_{3}^{z})$ \\
& 1 & 3 & $(0,0,0)$ & $J_{6}$ & $(D_{6}^{x},D_{6}^{y},D_{6}^{z})$ & $J_{3}$
& $(-D_{3}^{y},D_{3}^{x},-D_{3}^{z})$ \\
& 1 & 3 & $(0,-1,0)$ & $J_{6}$ & $(D_{6}^{x},D_{6}^{y},D_{6}^{z})$ & $J_{3}$
& $(-D_{3}^{y},D_{3}^{x},-D_{3}^{z})$ \\
& 2 & 4 & $(0,0,0)$ & $J_{6}$ & $(D_{6}^{x},D_{6}^{y},-D_{6}^{z})$ & $J_{3}$
& $(-D_{3}^{y},D_{3}^{x},D_{3}^{z})$ \\
& 2 & 4 & $(0,-1,0)$ & $J_{6}$ & $(D_{6}^{x},D_{6}^{y},-D_{6}^{z})$ & $J_{3}$
& $(-D_{3}^{y},D_{3}^{x},D_{3}^{z})$ \\
& 3 & 4 & $(0,0,0)$ & $J_{5}$ & $(D_{5}^{x},D_{5}^{y},-D_{5}^{z})$ & $J_{3}$
& $(D_{3}^{x},D_{3}^{y},-D_{3}^{z})$ \\
& 3 & 4 & $(-1,0,0)$ & $J_{5}$ & $(D_{5}^{x},D_{5}^{y},-D_{5}^{z})$ & $J_{3}$
& $(D_{3}^{x},D_{3}^{y},-D_{3}^{z})$ \\ \hline\hline
\end{tabular}%
\end{table*}

According to the above equations (i.e. Eq. (\ref{Jrelation-2}) and (\ref%
{Jrelation-3})), one can obtain the symmetry restricted MEIs for any space
group. Similarly, for the magnetic space group, the symmetry
restriction on MEIs can also be easily obtained. The collinear ferromagnetic
system shown in maintext (i.e. BNS 85.59 case) have two generators: the
four-fold rotation $\{4_{001}^{+}|1/2,0,0\}$ and inversion operation $\{%
\overline{1}|0,0,0\}$. The magnetic ion located at $\tau _{1}$ position has
only two nearest neighbors, i.e. (${\boldsymbol{\tau }_{1},\boldsymbol{\tau }%
_{5}+\mathbf{R}_{-100}}$) pair and (${\boldsymbol{\tau }_{1},\boldsymbol{%
\tau }_{6}+\mathbf{R}_{0-10}}$) pair as shown in Table \ref{non1} (also see
the Table \ref{1NN} in the maintext). These two bonds are equivalent by the
inversion symmetry $\{\overline{1}|0,0,0\}$. Meanwhile, performing the
four-fold rotation symmetry for the above two pairs, we can get other six
pairs in a unit cell, and there are in total eight NN in a unit cell. Based
on Eq. (\ref{Jrelation-2}), it is also easy to prove all of these eight NN
exchange paths has the same Heisenberg term, which we denote as $J_{1}$ as
shown in Table \ref{non1}. The non-collinearity shown in maintext reduces
the four-fold rotation $\{4_{001}^{+}|1/2,0,0\}$ to the two-fold rotation
operation $\{2_{001}|1/2,1/2,0\}$, as a result the eight NN exchange path is
no longer equivalent as indicated in Table \ref{non1}. Similarly, one can
obtain the symmetry restriction on the DM interactions, as also shown in
Table \ref{non1}.

For simplicity, we only list the corresponding MEIs for longer range with
the collinear FM example (i.e. the case with symmetry of
BNS 85.59) and non-collinear example in the maintext (i.e. the case with symmetry of BNS
13.65), as shown in Table \ref{non1-2}.

\begin{table}[tbph]
\caption{The corresponding MEIs for the collinear FM example (i.e. the case with symmetry of BNS 85.59) and non-collinear example (i.e.
the case with symmetry of BNS 13.65) shown in maintext.}
\label{non1-2}\centering%
\begin{tabular}{ll}
\hline\hline
BNS 85.59 & BNS 13.65 \\ \hline
$J_{1}$ & $J_{1},J_{2}$ \\
$J_{2}$ & $J_{3},J_{4}$ \\
$J_{3}$ & $J_{5},J_{6}$ \\
$J_{4}$ & $J_{7},J_{8}$ \\
$J_{5}\symbol{126}J_{7}$ & $J_{9}\symbol{126}J_{14}$ \\
$J_{8}$ & $J_{15},J_{16}$ \\
$J_{9},J_{10}$ & $J_{17}\symbol{126}J_{20}$ \\
$J_{11},J_{12}$ & $J_{21}\symbol{126}J_{24}$ \\
$J_{13}\symbol{126}J_{15}$ & $J_{25}\symbol{126}J_{28}$ \\
$J_{16}$ & $J_{29},J_{30}$ \\
$J_{17},J_{18}$ & $J_{32}\symbol{126}J_{34}$ \\
$J_{19}$ & $J_{35},J_{36}$ \\
$J_{20},J_{21}$ & $J_{37}\symbol{126}J_{40}$ \\
$J_{22}$ & $J_{41},J_{42}$ \\
$J_{23}$ & $J_{43},J_{44}$ \\
$J_{24}\symbol{126}J_{27}$ & $J_{45}\symbol{126}J_{52}$ \\ \hline\hline
\end{tabular}%
\end{table}

It is worth mentioning that Eq. (\ref{Jrelation}) can also give symmetry
restrictions on SIA. The general quadratic expression of SIA could be
written as $\sum_{l,n,\alpha ,\beta }K_{\mathbf{R}_{l}+\boldsymbol{\tau }%
_{n}}^{\alpha ,\beta }S_{ln}^{\alpha }S_{ln}^{\beta }$. Note that the SIA
term should be naturally symmetric, i.e. $K_{\mathbf{R}_{l}+\boldsymbol{\tau
}_{n}}^{\alpha ,\beta }=K_{\mathbf{R}_{l}+\boldsymbol{\tau }_{n}}^{\beta
,\alpha }$. According to Eq. (\ref{Jrelation}), when the action of symmetry
operation $\{\alpha |t\}$ keeps the position $R_{l}+\tau _{n}$ unchanged, we
have $K_{\mathbf{R}_{l}+\boldsymbol{\tau }_{n}}=M(\alpha )K_{\mathbf{R}_{l}+%
\boldsymbol{\tau }_{n}}M^{\dag }(\alpha )$. For the typical FM collinear
magnetic material (BNS 85.59) in the maintext, the SIA term for magnetic
ions located at $2a$ and $2c$ Wyckoff positions should satisfy that

\begin{equation*}
K_{\mathbf{R}_{l}+\boldsymbol{\tau }_{n}}=\left[
\begin{array}{ccc}
K_{\mathbf{R}_{l}+\boldsymbol{\tau }_{n}}^{xx} &  &  \\
& K_{\mathbf{R}_{l}+\boldsymbol{\tau }_{n}}^{xx} &  \\
&  & K_{\mathbf{R}_{l}+\boldsymbol{\tau }_{n}}^{zz}%
\end{array}%
\right]
\end{equation*}

which is actually the usual form $H_{SIA}=\sum_{l,n}K_{\mathbf{R}_{l}+%
\boldsymbol{\tau }_{n}}(S_{ln}^{z})^{2}+C$ where $K_{\mathbf{R}_{l}+%
\boldsymbol{\tau }_{n}}=K_{\mathbf{R}_{l}+\boldsymbol{\tau }_{n}}^{zz}-K_{%
\mathbf{R}_{l}+\boldsymbol{\tau }_{n}}^{xx}$ and $C$ represents the constant
term. However, the symmetry does not give restrictions on the SIA term of
magnetic ions at $4d$ position. In the maintext, we adopt the usual form $%
H_{SIA}=\sum_{l,n}K(S_{ln}^{z})^{2}$ \cite{book-3} for simplicity.

\subsection{The parameters $A_{n,n^{\prime }}$, $B_{n,n^{\prime }}$, $%
C_{n,n^{\prime }}$, $\mathbf{O}_{n,n^{\prime }}$, $\mathbf{P}_{n,n^{\prime
}} $ and $\mathbf{Q}_{n,n^{\prime }}$}

In the maintext, considering the spin model including the Heisenberg and DM
interactions, we perform the standard LSWT and obtain spin Hamiltonian as
Eq. \ref{ourHk}, where the parameters $A_{n,n^{\prime }}$, $B_{n,n^{\prime
}} $, $C_{n,n^{\prime }}$, $\mathbf{O}_{n,n^{\prime }}$, $\mathbf{P}%
_{n,n^{\prime }}$ and $\mathbf{Q}_{n,n^{\prime }}$ are related to the spin
directions at $n$ and $n^{\prime }$ sites. Here $A_{n,n^{\prime }}$, $%
B_{n,n^{\prime }}$ and $C_{n,n^{\prime }}$ could be written as:

\begin{eqnarray}
A_{n,n^{\prime }} &=&\frac{1}{4}\cos (\theta _{n}-\theta _{n^{\prime }})-%
\frac{1}{4}\cos (\theta _{n}+\theta _{n^{\prime }})+\frac{1}{2}\cos (\phi
_{n}-\phi _{n^{\prime }})  \notag \\
&&+\frac{1}{8}\cos (\theta _{n}-\theta _{n^{\prime }}+\phi _{n}-\phi
_{n^{\prime }})  \notag \\
&&+\frac{1}{8}\cos (\theta _{n}+\theta _{n^{\prime }}+\phi _{n}-\phi
_{n^{\prime }})  \notag \\
&&+\frac{1}{8}\cos (\theta _{n}-\theta _{n^{\prime }}-\phi _{n}+\phi
_{n^{\prime }})  \notag \\
&&+\frac{1}{8}\cos (\theta _{n}+\theta _{n^{\prime }}-\phi _{n}+\phi
_{n^{\prime }})  \notag \\
&&+\frac{i}{4}\sin (\theta _{n}+\phi _{n}-\phi _{n^{\prime }})+\frac{i}{4}%
\sin (\theta _{n^{\prime }}+\phi _{n}-\phi _{n^{\prime }})  \notag \\
&&-\frac{i}{4}\sin (\theta _{n}-\phi _{n}+\phi _{n^{\prime }})-\frac{i}{4}%
\sin (\theta _{n^{\prime }}-\phi _{n}+\phi _{n^{\prime }}) \\
B_{n,n^{\prime }} &=&\sin \theta _{n}\sin \theta _{n^{\prime }}\cos (\phi
_{n}-\phi _{n^{\prime }})+\cos \theta _{n}\cos \theta _{n^{\prime }} \\
C_{n,n^{\prime }} &=&\frac{1}{4}\cos (\theta _{n}-\theta _{n^{\prime }})-%
\frac{1}{4}\cos (\theta _{n}+\theta _{n^{\prime }})-\frac{1}{2}\cos (\phi
_{n}-\phi _{n^{\prime }})  \notag \\
&&+\frac{1}{8}\cos (\theta _{n}-\theta _{n^{\prime }}+\phi _{n}-\phi
_{n^{\prime }})  \notag \\
&&+\frac{1}{8}\cos (\theta _{n}+\theta _{n^{\prime }}+\phi _{n}-\phi
_{n^{\prime }})  \notag \\
&&+\frac{1}{8}\cos (\theta _{n}-\theta _{n^{\prime }}-\phi _{n}+\phi
_{n^{\prime }})  \notag \\
&&+\frac{1}{8}\cos (\theta _{n}+\theta _{n^{\prime }}-\phi _{n}+\phi
_{n^{\prime }})  \notag \\
&&-\frac{i}{4}\sin (\theta _{n}+\phi _{n}-\phi _{n^{\prime }})+\frac{i}{4}%
\sin (\theta _{n^{\prime }}+\phi _{n}-\phi _{n^{\prime }})  \notag \\
&&+\frac{i}{4}\sin (\theta _{n}-\phi _{n}+\phi _{n^{\prime }})-\frac{i}{4}%
\sin (\theta _{n^{\prime }}-\phi _{n}+\phi _{n^{\prime }})
\end{eqnarray}

Meanwhile, the parameters $\mathbf{O}_{n,n^{\prime }}=(O_{n,n^{\prime
}}^{x},O_{n,n^{\prime }}^{y},O_{n,n^{\prime }}^{z})$, $\mathbf{P}%
_{n,n^{\prime }}=(P_{n,n^{\prime }}^{x},P_{n,n^{\prime }}^{y},P_{n,n^{\prime
}}^{z})$ and $\mathbf{Q}_{n,n^{\prime }}=(Q_{n,n^{\prime
}}^{x},Q_{n,n^{\prime }}^{y},Q_{n,n^{\prime }}^{z})$ related to the spin
directions at $n$ and $n^{\prime }$ sites could be written as:

\begin{eqnarray}
O_{n,n^{\prime }}^{x} &=&\frac{1}{2}[-\cos \theta _{n}\sin \theta
_{n^{\prime }}\sin \phi _{n}+\sin \theta _{n}\cos \theta _{n^{\prime }}\sin
\phi _{n^{\prime }}  \notag \\
&&+i(\sin \theta _{n}\cos \phi _{n^{\prime }}+\sin \theta _{n^{\prime }}\cos
\phi _{n})] \\
P_{n,n^{\prime }}^{x} &=&\sin \theta _{n}\cos \theta _{n^{\prime }}\sin \phi
_{n}-\cos \theta _{n}\sin \theta _{n^{\prime }}\sin \phi _{n^{\prime }} \\
Q_{n,n^{\prime }}^{x} &=&\frac{1}{2}[-\cos \theta _{n}\sin \theta
_{n^{\prime }}\sin \phi _{n}+\sin \theta _{n}\cos \theta _{n^{\prime }}\sin
\phi _{n^{\prime }}  \notag \\
&&+i(\sin \theta _{n}\cos \phi _{n^{\prime }}-\sin \theta _{n^{\prime }}\cos
\phi _{n})]
\end{eqnarray}

\begin{eqnarray}
O_{n,n^{\prime }}^{y} &=&\frac{1}{2}[-\cos \theta _{n}\sin \theta
_{n^{\prime }}\cos \phi _{n}+\sin \theta _{n}\cos \theta _{n^{\prime }}\cos
\phi _{n^{\prime }}  \notag \\
&&+i(-\sin \theta _{n}\sin \phi _{n^{\prime }}+\sin \theta _{n^{\prime
}}\sin \phi _{n})] \\
P_{n,n^{\prime }}^{y} &=&\sin \theta _{n}\cos \theta _{n^{\prime }}\cos \phi
_{n}-\cos \theta _{n}\sin \theta _{n^{\prime }}\cos \phi _{n^{\prime }} \\
Q_{n,n^{\prime }}^{y} &=&\frac{1}{2}[-\cos \theta _{n}\sin \theta
_{n^{\prime }}\cos \phi _{n}+\sin \theta _{n}\cos \theta _{n^{\prime }}\cos
\phi _{n^{\prime }}  \notag \\
&&+i(-\sin \theta _{n}\sin \phi _{n^{\prime }}-\sin \theta _{n^{\prime
}}\sin \phi _{n})]
\end{eqnarray}

\begin{eqnarray}
O_{n,n^{\prime }}^{z} &=&\frac{1}{2}[(1+\cos \theta _{n}\cos \theta
_{n^{\prime }})\sin (\phi _{n}-\phi _{n^{\prime }})  \notag \\
&&-i(\cos \theta _{n}+\cos \theta _{n^{\prime }})\cos (\phi _{n}-\phi
_{n^{\prime }})] \\
P_{n,n^{\prime }}^{z} &=&\sin \theta _{n}\sin \theta _{n^{\prime }}\sin
(\phi _{n}-\phi _{n^{\prime }}) \\
Q_{n,n^{\prime }}^{z} &=&\frac{1}{2}[(-1+\cos \theta _{n}\cos \theta
_{n^{\prime }})\sin (\phi _{n}-\phi _{n^{\prime }})  \notag \\
&&-i(\cos \theta _{n}-\cos \theta _{n^{\prime }})\cos (\phi _{n}-\phi
_{n^{\prime }})]
\end{eqnarray}

If the ground state of this magnetic system is collinear, which means that $%
\theta _{n}=\theta _{n^{\prime }}$ (when magnetic ions at $n$ and $n^{\prime
}$ sites are ferromagnetic) or $\theta _{n}=\theta _{n^{\prime }}+\pi $
(when magnetic ions at $n$ and $n^{\prime }$ sites are antiferromagnetic)
and $\phi _{n}=\phi _{n^{\prime }}$, the parameters $A_{n,n^{\prime }}$, $%
B_{n,n^{\prime }}$, $C_{n,n^{\prime }}$ can be simplified as

\begin{eqnarray}
A_{n,n^{\prime }} &=&\frac{\zeta _{n,n^{\prime }}+1}{2} \\
B_{n,n^{\prime }} &=&\zeta _{n,n^{\prime }} \\
C_{n,n^{\prime }} &=&\frac{\zeta _{n,n^{\prime }}-1}{2}
\end{eqnarray}

where $\zeta _{n,n^{\prime }}$ equals to 1 when the spins for the $n$ and $%
n^{\prime }$ sites are parallel, otherwise $\zeta _{n,n^{\prime }}$ equals
to $-1$. More specially, when the ground state of this magnetic system is
assumed to be collinear ferromagnetic [PS: do not need to be along
z-direction], which means that $\theta _{n}=\theta _{n^{\prime }}$ and $\phi
_{n}=\phi _{n^{\prime }}$, the parameters $A_{n,n^{\prime }}$, $%
B_{n,n^{\prime }}$, $C_{n,n^{\prime }}$ can be simplified as

\begin{eqnarray}
A_{n,n^{\prime }} &=&1  \label{fma} \\
B_{n,n^{\prime }} &=&1  \label{fmb} \\
C_{n,n^{\prime }} &=&0  \label{fmc}
\end{eqnarray}

\subsection{The eigenvalue problem of magnon eigenvalues}

As shown in the maintext, following the LSWT, a general pairwise spin
Hamiltonian could be written as Eq. (\ref{2}). However, it should be noted
that the operators in $\psi ^{\dag }(\mathbf{k})=[a_{1}^{\dag }(\mathbf{k}%
),...,a_{N}^{\dag }(\mathbf{k}),a_{1}(-\mathbf{k}),...,a_{N}(-\mathbf{k})]$
in the Eq. (3) of the maintext satisfy the commutation relation%
\begin{equation}
\lbrack \psi (k),\psi ^{\dag }(k)]=%
\begin{bmatrix}
I &  \\
& -I%
\end{bmatrix}%
=I_{-}  \label{i-}
\end{equation}

where $I$ represents N$\times $N identity matrix, and N represents the
number of magnetic ions in an unit cell. To diagonalize the boson pairing
Hamiltonian, we can solve the eigenvalue problem of the general Hamiltonian $%
H_{J}(\mathbf{k})=I_{-}H(\mathbf{k})$ (i.e., Eq. (6) of the maintext). The
first N diagonal elements are the energies of the normal spin wave modes $%
\omega {_{k}},{_{n}}$ and the last N eigenvalues are equal to the first N
eigenvalues multiplied by minus one.

\subsection{Instructions on the program of general relations between magnon
eigenvalues}

As shown in the maintext, a general pairwise spin model could be expanded as
the isotropic Heisenberg Hamiltonian, the DM interactions, and the
anisotropic symmetric terms, as shown in Eq. (2) in the maintext. We ignore
the third term and perform LSWT to obtain the quadratic spin Hamiltonian as
shown in Eq. (3) in the maintext. Note that there is a simple relation
between the Fourier transform of MEIs and the SSME, consequently one can
easily calculate SSME at arbitrary $k$ point in BZ. Thus, different with the
conventional group symmetry analysis, which give the relationship between
the magnon energies at the symmetry-related $k$ points, our method produces
the relationships between the SSME at high symmetry k points subjected to $%
R_{cut}$. For any given magnetic system, we propose a method to obtain the
relations between SSME of different high-symmetry k-points. The algorithm of
the proposed method is implemented in the Mathematica notebook "SR.nb".
Using this code, one should first enter the information of magnetic
materials, including: (1) the primitive basis and conventional basis; (2)
the positions of magnetic atoms; (3) the (magnetic) space group; (4) the
spin directions of magnetic atoms; (5) the range of MEIs to be considered
(including Heisenberg and DM interactions); (6) whether to consider SIA; (7)
the interested k-points. Then the relations between SSME can be
automatically obtained. To catch your eye, we have used red color to
indicate that the following variable should be specified in the notebook
"SR.nb". Below, we present an example for the typical magnetic system (BNS
85.59).

As shown in Table \ref{positions} in the maintext, the lattice constant $c/a$
is 0.8. The magnetic ions are located at three nonequivalent
crystallographic sites: $4d$ (0, 0, 0), $2a$ (0.25, 0.75, 0) and $2c$ (0.25,
0.25, $z$) WP and the positions for these eight magnetic ions are summarized
in Table \ref{positions} in the maintext. While the $4d$ and $2a$ WP had
been completely determined by the symmetry, the coordinates of $2c$ WP have
a variable $z$ and here we adopt it as $z=0.1$. The magnetic state is a
collinear ferromagnetic order with spin along the $z$ direction. This case
belongs to the type-I magnetic space group (BNS 85.59), and all of the the
polar angle and azimuthal angles are equal to 0.

In this notebook "SR.nb", one should specify the parameters as input
information, such as:

\bigskip

\textcolor{red}{
(*input parameters:*)
}

\textcolor{red}{
(*primitive lattice basis*)
}

A = \{\{1, 0, 0\}, \{0, 1, 0\}, \{0, 0, 0.8\}\};

\textcolor{red}{
(*conventional lattice basis*)
}

AA = \{\{1, 0, 0\}, \{0, 1, 0\}, \{0, 0, 0.8\}\};

\textcolor{red}{
(*input positions of magnetic atoms based on conventional lattice basis vectors*)
}

atoms=8;

$\boldsymbol{\tau }=$ \{\{0, 0, 0\},

\{0.5, 0, 0\},

\{0, 0.5, 0\},

\{0.5, 0.5, 0\},

\{0.75, 0.25, 0.\},

\{0.25, 0.75, 0.\},

\{0.25, 0.25, 0.1\},

\{0.75, 0.75, -0.1\}\};

\textcolor{red}{
(*input the serial number of space group or the magnetic space group (in BNS notation).*)
}

msg=85.59;

\textcolor{red}{
(*spin directions of magnetic atoms in spherical coordinates ($\theta _{n},\phi _{n}$) with the polar angles $\theta _{n}$\ and azimuthal angles $\phi _{n}$.*)
}

ang = \{\{0, 0\}, \{0, 0\}, \{0, 0\}, \{0, 0\}, \{0, 0\}, \{0, 0\}, \{0,
0\}, \{0,0\}\};

\textcolor{red}{
(*the range of Heisenberg and DM interactions to be considered, respectively*)
}

Jmax = 3;

Dmax = 2;

\textcolor{red}{
(*set DIA = "True" or "False" to indicate whether to consider single ion anisotropy.*)
}

DIA = False;

\textcolor{red}{
(*high-symmetry points*)
}

kk[1] = \{0, 0, 0\};

kk[2] = \{0.5, 0, 0\};

kk[3] = \{0.5, 0.5, 0\};

kk[4] = \{0, 0, 0.5\};

kk[5] = \{0.5, 0, 0.5\};

kk[6] = \{0.5, 0.5, 0.5\};

kname = \{"$\Gamma $", "X", "M", "Z", "R", "A"\};

\bigskip

In the following, we would like to give a description of these parameters
one by one:

\bigskip

(1) the primitive basis and conventional basis of Bravais lattice

One should input them in Cartesian coordinates. As the example in the
maintext, the space group P4/n (No. 85) crystallizes in a tetragonal
lattice, and its primitive basis $A$ and conventional basis $AA$ of Bravais
lattice are both $\{\{1,0,0\},\{0,1,0\},\{0,0,0.8\}\}$.

\textcolor{red}{
(*primitive lattice basis*)
}

A = \{\{1, 0, 0\}, \{0, 1, 0\}, \{0, 0, 0.8\}\};

\textcolor{red}{
(*conventional lattice basis*)
}

AA = \{\{1, 0, 0\}, \{0, 1, 0\}, \{0, 0, 0.8\}\};

\bigskip (2) the number and the Wyckoff positions of magnetic atoms (based
on conventional lattice basis vectors)

\textcolor{red}{
(*input positions of magnetic atoms based on conventional lattice basis vectors*)
}

atoms=8;

$\boldsymbol{\tau }=$ \{\{0, 0, 0\},

\{0.5, 0, 0\},

\{0, 0.5, 0\},

\{0.5, 0.5, 0\},

\{0.75, 0.25, 0.\},

\{0.25, 0.75, 0.\},

\{0.25, 0.25, 0.1\},

\{0.75, 0.75, -0.1\}\};

\bigskip

(3) the (magnetic) space group of the magnetic system. When the magnetic
moments are quite localized, magnetic interactions may still satisfy the
symmetries of its space group. In this case, one can enter the serial number
of its space group. Otherwise, one should enter the serial number of its
magnetic space group (in BNS notation). As the example for collinear
ferromagnetic state in the maintext, the magnetic space group (BNS 85.59)
has the same symmetries as the space group (SG. 85). In this case, the
results of the input "msg=85" and "msg=85.59" are equivalent. Note that if
the input parameter is an integer "X", we would use the symmetry of this
space group (No. X), otherwise, we will use the symmetry of the magnetic
space group as "BNS\ X.Y".

\textcolor{red}{
(*input the serial number of space group or the magnetic space group (in BNS notation).*)
}

msg=85.59;

\bigskip (4) the spin directions of magnetic atoms. One should input them in
spherical coordinates ($\theta _{n},\phi _{n}$) with the polar angles $%
\theta _{n}$\ and azimuthal angles $\phi _{n}$. As the example in the
maintext, the FM (001) state should be

\textcolor{red}{
(*spin directions of magnetic atoms in spherical coordinates ($\theta _{n},\phi _{n}$) with the polar angles $\theta _{n}$\ and azimuthal angles $\phi _{n}$.*)
}

ang = \{\{0, 0\}, \{0, 0\}, \{0, 0\}, \{0, 0\}, \{0, 0\}, \{0, 0\}, \{0,
0\}, \{0,0\}\};

\bigskip

(5) One should set up the range of magnetic interactions to be considered.
You can set the range of Heisenberg and DM interactions respectively.

\textcolor{red}{
(*the range of Heisenberg and DM interactions to be considered, respectively*)
}

Jmax = 3;

Dmax = 2;

For example, we set Jmax = 3 and Dmax = 2, which means that we consider the
range of Heisenberg interactions up to $J_{3}$, while the DM interactions
are considered up to $\mathbf{D}_{2}$.

\bigskip

(6) Then one should set up whether to consider SIA by setting DIA = "True"
or "False".

\textcolor{red}{
(*set DIA = "True" or "False" to indicate whether to consider single ion anisotropy.*)
}

DIA = False;

\bigskip

(7) Last, one should input the information of high-symmetry points to be
considered, including the positions of these high-symmetry points and the
labeled names of these high-symmetry points.

\textcolor{red}{
(*high-symmetry points*)
}

kk[1] = \{0, 0, 0\};

kk[2] = \{0.5, 0, 0\};

kk[3] = \{0.5, 0.5, 0\};

kk[4] = \{0, 0, 0.5\};

kk[5] = \{0.5, 0, 0.5\};

kk[6] = \{0.5, 0.5, 0.5\};

kname = \{"$\Gamma $", "X", "M", "Z", "R", "A"\};

\bigskip

After entering the above parameters, our program would output the following
information, including:

\begin{table}[h]
\caption{The Heisenberg and DM interactions for the NNs of the typical
material in the maintext for collinear FM state (BNS 85.59) restricted by
the crystal symmetry. }
\label{1dm}\centering%
\begin{tabular}{cccccc}
\hline\hline
distance(a) & $n$ & $n^{\prime }$ & $R_{l}$ & $J$ & D \\ \hline
0.35 & 1 & 5 & $(-1,0,0)$ & $J_{1}$ & $(D_{1}^{x},D_{1}^{y},D_{1}^{z})$ \\
& 1 & 6 & $(0,-1,0)$ &  & $(D_{1}^{x},D_{1}^{y},D_{1}^{z})$ \\
& 2 & 5 & $(0,0,0)$ &  & $(-D_{1}^{y},D_{1}^{x},D_{1}^{z})$ \\
& 2 & 6 & $(0,-1,0)$ &  & $(-D_{1}^{y},D_{1}^{x},D_{1}^{z})$ \\
& 3 & 5 & $(-1,0,0)$ &  & $(D_{1}^{y},-D_{1}^{x},D_{1}^{z})$ \\
& 3 & 6 & $(0,0,0)$ &  & $(D_{1}^{y},-D_{1}^{x},D_{1}^{z})$ \\
& 4 & 5 & $(0,0,0)$ &  & $(-D_{1}^{x},-D_{1}^{y},D_{1}^{z})$ \\
& 4 & 6 & $(0,0,0)$ &  & $(-D_{1}^{x},-D_{1}^{y},D_{1}^{z})$ \\ \hline
0.36 & 1 & 7 & $(0,0,0)$ & $J_{2}$ & $(D_{2}^{x},D_{2}^{y},D_{2}^{z})$ \\
& 1 & 8 & $(-1,-1,0)$ &  & $(D_{2}^{x},D_{2}^{y},D_{2}^{z})$ \\
& 2 & 7 & $(0,0,0)$ &  & $(-D_{2}^{y},D_{2}^{x},D_{2}^{z})$ \\
& 2 & 8 & $(0,-1,0)$ &  & $(-D_{2}^{y},D_{2}^{x},D_{2}^{z})$ \\
& 3 & 7 & $(0,0,0)$ &  & $(D_{2}^{y},-D_{2}^{x},D_{2}^{z})$ \\
& 3 & 8 & $(-1,0,0)$ &  & $(D_{2}^{y},-D_{2}^{x},D_{2}^{z})$ \\
& 4 & 7 & $(0,0,0)$ &  & $(-D_{2}^{x},-D_{2}^{y},D_{2}^{z})$ \\
& 4 & 8 & $(0,0,0)$ &  & $(-D_{2}^{x},-D_{2}^{y},D_{2}^{z})$ \\ \hline
0.5 & 1 & 2 & $(0,0,0)$ & $J_{3}$ & $(D_{3}^{x},D_{3}^{y},D_{3}^{z})$ \\
& 1 & 2 & $(-1,0,0)$ &  & $(D_{3}^{x},D_{3}^{y},D_{3}^{z})$ \\
& 1 & 3 & $(0,0,0)$ &  & $(-D_{3}^{y},D_{3}^{x},-D_{3}^{z})$ \\
& 1 & 3 & $(0,-1,0)$ &  & $(-D_{3}^{y},D_{3}^{x},-D_{3}^{z})$ \\
& 2 & 4 & $(0,0,0)$ &  & $(-D_{3}^{y},D_{3}^{x},D_{3}^{z})$ \\
& 2 & 4 & $(0,-1,0)$ &  & $(-D_{3}^{y},D_{3}^{x},D_{3}^{z})$ \\
& 3 & 4 & $(0,0,0)$ &  & $(D_{3}^{x},D_{3}^{y},-D_{3}^{z})$ \\
& 3 & 4 & $(-1,0,0)$ &  & $(D_{3}^{x},D_{3}^{y},-D_{3}^{z})$ \\ \hline
\end{tabular}%
\end{table}

\bigskip

(1) the symmetry of the (magnetic) space group.

output[symmetry]=

(x,y,z $\mid $ mx,my,mz)

(-y+1/2,x,z $\mid$ -my,mx,mz)

(y,-x+1/2,z $\mid$ my,-mx,mz)

(-x+1/2,-y+1/2,z $\mid$ -mx,-my,mz)

(-x,-y,-z $\mid$ mx,my,mz)

(y+1/2,-x,-z $\mid$ -my,mx,mz)

(-y,x+1/2,-z $\mid$ my,-mx,mz)

(x+1/2,y+1/2,-z $\mid $ -mx,-my,mz)

where the left part represents the symmetry\ operation for positions of
magnetic atoms, while the right part means the symmetry operation for the
orientation of magnetic moment.

\bigskip

(2) The program would also give the distance and the corresponding symmetry
restricted\ MEIs, including the Heisenberg and DM interactions. For example,
there are in total 24 MEIs in this magnetic system up to $J_{3}$, as
summarized in the Table \ref{1dm}. Meanwhile, the corresponding symmetry
restricted\ DM interactions $\mathbf{D}(\boldsymbol{\tau }_{n},\boldsymbol{%
\tau }_{n^{\prime }},\mathbf{R}_{l})$ are also listed here. Note that for
5th NN MEIs, the symmetry makes $D_{5}^{z}=0$, while for 7th NN MEIs, we
have $\mathbf{D}_{5}=(0,0,0)$. These symmetry restrictions would also be automatically considered in our program.

\bigskip

(3) the main output: the relations between SSME at different high-symmetry $%
k $ points.

output[relations]=

$M-A=0$

$X-R=0$

$\Gamma -Z=0$

$\Gamma -2X+M{}=0$

where the label "M" means the quadratic sum of the magnon energies at M
point "$\sum_{i}\omega _{i}^{2}(M){}$", as well as the labels of other
high-symmetry $k$ points. We can see that, up to $J_{5}$, the quadratic sum
of the magnon energies satisfy that $\sum_{i}\omega _{i}^{2}(\Gamma
)=\sum_{i}\omega _{i}^{2}(Z)$, $\sum_{i}\omega _{i}^{2}(X)=\sum_{i}\omega
_{i}^{2}(R)$, $\sum_{i}\omega _{i}^{2}(M)=\sum_{i}\omega _{i}^{2}(A)$ and $%
2\sum_{i}\omega _{i}^{2}(X){}=\sum_{i}\omega _{i}^{2}(\Gamma
)+\sum_{i}\omega _{i}^{2}(M)$, as shown in Table \ref{relation1} in the
maintext.

\bigskip

As shown above, by entering the information of magnetic materials, one can
use this code to obtain the relations between magnon eigenvalues easily.

\bibliographystyle{aps}
\bibliography{sumrule2}

\providecommand{\noopsort}[1]{}\providecommand{\singleletter}[1]{#1}%
\begin{thebibliography}{47}%
\makeatletter
\providecommand \@ifxundefined [1]{%
 \@ifx{#1\undefined}
}%
\providecommand \@ifnum [1]{%
 \ifnum #1\expandafter \@firstoftwo
 \else \expandafter \@secondoftwo
 \fi
}%
\providecommand \@ifx [1]{%
 \ifx #1\expandafter \@firstoftwo
 \else \expandafter \@secondoftwo
 \fi
}%
\providecommand \natexlab [1]{#1}%
\providecommand \enquote  [1]{``#1''}%
\providecommand \bibnamefont  [1]{#1}%
\providecommand \bibfnamefont [1]{#1}%
\providecommand \citenamefont [1]{#1}%
\providecommand \href@noop [0]{\@secondoftwo}%
\providecommand \href [0]{\begingroup \@sanitize@url \@href}%
\providecommand \@href[1]{\@@startlink{#1}\@@href}%
\providecommand \@@href[1]{\endgroup#1\@@endlink}%
\providecommand \@sanitize@url [0]{\catcode `\\12\catcode `\$12\catcode
  `\&12\catcode `\#12\catcode `\^12\catcode `\_12\catcode `\%12\relax}%
\providecommand \@@startlink[1]{}%
\providecommand \@@endlink[0]{}%
\providecommand \url  [0]{\begingroup\@sanitize@url \@url }%
\providecommand \@url [1]{\endgroup\@href {#1}{\urlprefix }}%
\providecommand \urlprefix  [0]{URL }%
\providecommand \Eprint [0]{\href }%
\providecommand \doibase [0]{http://dx.doi.org/}%
\providecommand \selectlanguage [0]{\@gobble}%
\providecommand \bibinfo  [0]{\@secondoftwo}%
\providecommand \bibfield  [0]{\@secondoftwo}%
\providecommand \translation [1]{[#1]}%
\providecommand \BibitemOpen [0]{}%
\providecommand \bibitemStop [0]{}%
\providecommand \bibitemNoStop [0]{.\EOS\space}%
\providecommand \EOS [0]{\spacefactor3000\relax}%
\providecommand \BibitemShut  [1]{\csname bibitem#1\endcsname}%
\let\auto@bib@innerbib\@empty
\bibitem [{\citenamefont {St{\"o}hr}\ and\ \citenamefont
  {Siegmann}(2006)}]{book-1}%
  \BibitemOpen
  \bibfield  {author} {\bibinfo {author} {\bibfnamefont {J.}~\bibnamefont
  {St{\"o}hr}}\ and\ \bibinfo {author} {\bibfnamefont {H.}~\bibnamefont
  {Siegmann}},\ }\href@noop {} {\emph {\bibinfo {title} {Magnetism From
  Fundamentals to Nanoscale Dynamics}}}\ (\bibinfo  {publisher} {Springer},\
  \bibinfo {year} {2006})\BibitemShut {NoStop}%
\bibitem [{\citenamefont {Buschow}\ \emph {et~al.}(2003)\citenamefont
  {Buschow}, \citenamefont {Boer} \emph {et~al.}}]{book-2}%
  \BibitemOpen
  \bibfield  {author} {\bibinfo {author} {\bibfnamefont {K.~H.~J.}\
  \bibnamefont {Buschow}}, \bibinfo {author} {\bibfnamefont {F.~R.}\
  \bibnamefont {Boer}},  \emph {et~al.},\ }\href@noop {} {\emph {\bibinfo
  {title} {Physics of Magnetism and Magnetic Materials}}}\ (\bibinfo
  {publisher} {Springer},\ \bibinfo {year} {2003})\BibitemShut {NoStop}%
\bibitem [{\citenamefont {White}(2007)}]{book-3}%
  \BibitemOpen
  \bibfield  {author} {\bibinfo {author} {\bibfnamefont {R.~M.}\ \bibnamefont
  {White}},\ }\href@noop {} {\emph {\bibinfo {title} {Quantum Theory of
  Magnetism: Magnetic Properties of Materials}}}\ (\bibinfo  {publisher}
  {Springer-Verlag Berlin Heidelberg},\ \bibinfo {year} {2007})\BibitemShut
  {NoStop}%
\bibitem [{\citenamefont {Lichtenstein}\ \emph {et~al.}(2003)\citenamefont
  {Lichtenstein}, \citenamefont {Anisimov},\ and\ \citenamefont
  {Katsnelson}}]{Lichtenstein-book}%
  \BibitemOpen
  \bibfield  {author} {\bibinfo {author} {\bibfnamefont {A.~I.}\ \bibnamefont
  {Lichtenstein}}, \bibinfo {author} {\bibfnamefont {V.~I.}\ \bibnamefont
  {Anisimov}}, \ and\ \bibinfo {author} {\bibfnamefont {M.~I.}\ \bibnamefont
  {Katsnelson}},\ }\href@noop {} {\emph {\bibinfo {title} {Electronic Structure
  and Magnetism of Correlated Systems: Beyond LDA}}}\ (\bibinfo  {publisher}
  {Springer},\ \bibinfo {year} {2003})\BibitemShut {NoStop}%
\bibitem [{\citenamefont {Prabhakar}\ and\ \citenamefont
  {Stancil}(2009)}]{spinwavebook}%
  \BibitemOpen
  \bibfield  {author} {\bibinfo {author} {\bibfnamefont {A.}~\bibnamefont
  {Prabhakar}}\ and\ \bibinfo {author} {\bibfnamefont {D.~D.}\ \bibnamefont
  {Stancil}},\ }\href@noop {} {\emph {\bibinfo {title} {Spin waves: Theory and
  applications}}},\ Vol.~\bibinfo {volume} {5}\ (\bibinfo  {publisher}
  {Springer},\ \bibinfo {year} {2009})\BibitemShut {NoStop}%
\bibitem [{\citenamefont {Krawczyk}\ and\ \citenamefont
  {Grundler}(2014)}]{spin-wave-1}%
  \BibitemOpen
  \bibfield  {author} {\bibinfo {author} {\bibfnamefont {M.}~\bibnamefont
  {Krawczyk}}\ and\ \bibinfo {author} {\bibfnamefont {D.}~\bibnamefont
  {Grundler}},\ }\bibfield  {title} {\enquote {\bibinfo {title} {Review and
  prospects of magnonic crystals and devices with reprogrammable band
  structure},}\ }\href@noop {} {\bibfield  {journal} {\bibinfo  {journal} {J.
  Phys.: Condens. Matter}\ }\textbf {\bibinfo {volume} {26}},\ \bibinfo {pages}
  {123202} (\bibinfo {year} {2014})}\BibitemShut {NoStop}%
\bibitem [{\citenamefont {Kosevich}\ \emph {et~al.}(1990)\citenamefont
  {Kosevich}, \citenamefont {Ivanov},\ and\ \citenamefont
  {Kovalev}}]{kosevich1990magnetic}%
  \BibitemOpen
  \bibfield  {author} {\bibinfo {author} {\bibfnamefont {A.~M.}\ \bibnamefont
  {Kosevich}}, \bibinfo {author} {\bibfnamefont {B.}~\bibnamefont {Ivanov}}, \
  and\ \bibinfo {author} {\bibfnamefont {A.}~\bibnamefont {Kovalev}},\
  }\bibfield  {title} {\enquote {\bibinfo {title} {Magnetic solitons},}\
  }\href@noop {} {\bibfield  {journal} {\bibinfo  {journal} {Physics Reports}\
  }\textbf {\bibinfo {volume} {194}},\ \bibinfo {pages} {117} (\bibinfo {year}
  {1990})}\BibitemShut {NoStop}%
\bibitem [{\citenamefont {Fogedby}(1980)}]{fogedby1980solitons}%
  \BibitemOpen
  \bibfield  {author} {\bibinfo {author} {\bibfnamefont {H.~C.}\ \bibnamefont
  {Fogedby}},\ }\bibfield  {title} {\enquote {\bibinfo {title} {Solitons and
  magnons in the classical heisenberg chain},}\ }\href@noop {} {\bibfield
  {journal} {\bibinfo  {journal} {Journal of Physics A: Mathematical and
  General}\ }\textbf {\bibinfo {volume} {13}},\ \bibinfo {pages} {1467}
  (\bibinfo {year} {1980})}\BibitemShut {NoStop}%
\bibitem [{\citenamefont {Giamarchi}\ \emph {et~al.}(2008)\citenamefont
  {Giamarchi}, \citenamefont {R{\"u}egg},\ and\ \citenamefont
  {Tchernyshyov}}]{giamarchi2008bose}%
  \BibitemOpen
  \bibfield  {author} {\bibinfo {author} {\bibfnamefont {T.}~\bibnamefont
  {Giamarchi}}, \bibinfo {author} {\bibfnamefont {C.}~\bibnamefont
  {R{\"u}egg}}, \ and\ \bibinfo {author} {\bibfnamefont {O.}~\bibnamefont
  {Tchernyshyov}},\ }\bibfield  {title} {\enquote {\bibinfo {title}
  {Bose--einstein condensation in magnetic insulators},}\ }\href@noop {}
  {\bibfield  {journal} {\bibinfo  {journal} {Nature Physics}\ }\textbf
  {\bibinfo {volume} {4}},\ \bibinfo {pages} {198} (\bibinfo {year}
  {2008})}\BibitemShut {NoStop}%
\bibitem [{\citenamefont {Nikuni}\ \emph {et~al.}(2000)\citenamefont {Nikuni},
  \citenamefont {Oshikawa}, \citenamefont {Oosawa},\ and\ \citenamefont
  {Tanaka}}]{nikuni2000bose}%
  \BibitemOpen
  \bibfield  {author} {\bibinfo {author} {\bibfnamefont {T.}~\bibnamefont
  {Nikuni}}, \bibinfo {author} {\bibfnamefont {M.}~\bibnamefont {Oshikawa}},
  \bibinfo {author} {\bibfnamefont {A.}~\bibnamefont {Oosawa}}, \ and\ \bibinfo
  {author} {\bibfnamefont {H.}~\bibnamefont {Tanaka}},\ }\bibfield  {title}
  {\enquote {\bibinfo {title} {Bose-einstein condensation of dilute magnons in
  tlcucl 3},}\ }\href@noop {} {\bibfield  {journal} {\bibinfo  {journal} {Phys.
  Rev. Lett.}\ }\textbf {\bibinfo {volume} {84}},\ \bibinfo {pages} {5868}
  (\bibinfo {year} {2000})}\BibitemShut {NoStop}%
\bibitem [{\citenamefont {Demokritov}\ \emph {et~al.}(2006)\citenamefont
  {Demokritov}, \citenamefont {Demidov}, \citenamefont {Dzyapko}, \citenamefont
  {Melkov}, \citenamefont {Serga}, \citenamefont {Hillebrands},\ and\
  \citenamefont {Slavin}}]{demokritov2006bose}%
  \BibitemOpen
  \bibfield  {author} {\bibinfo {author} {\bibfnamefont {S.}~\bibnamefont
  {Demokritov}}, \bibinfo {author} {\bibfnamefont {V.}~\bibnamefont {Demidov}},
  \bibinfo {author} {\bibfnamefont {O.}~\bibnamefont {Dzyapko}}, \bibinfo
  {author} {\bibfnamefont {G.}~\bibnamefont {Melkov}}, \bibinfo {author}
  {\bibfnamefont {A.}~\bibnamefont {Serga}}, \bibinfo {author} {\bibfnamefont
  {B.}~\bibnamefont {Hillebrands}}, \ and\ \bibinfo {author} {\bibfnamefont
  {A.}~\bibnamefont {Slavin}},\ }\bibfield  {title} {\enquote {\bibinfo {title}
  {Bose--einstein condensation of quasi-equilibrium magnons at room temperature
  under pumping},}\ }\href@noop {} {\bibfield  {journal} {\bibinfo  {journal}
  {Nature}\ }\textbf {\bibinfo {volume} {443}},\ \bibinfo {pages} {430}
  (\bibinfo {year} {2006})}\BibitemShut {NoStop}%
\bibitem [{\citenamefont {Onose}\ \emph {et~al.}(2010)\citenamefont {Onose},
  \citenamefont {Ideue}, \citenamefont {Katsura}, \citenamefont {Shiomi},
  \citenamefont {Nagaosa},\ and\ \citenamefont
  {Tokura}}]{onose2010observation}%
  \BibitemOpen
  \bibfield  {author} {\bibinfo {author} {\bibfnamefont {Y.}~\bibnamefont
  {Onose}}, \bibinfo {author} {\bibfnamefont {T.}~\bibnamefont {Ideue}},
  \bibinfo {author} {\bibfnamefont {H.}~\bibnamefont {Katsura}}, \bibinfo
  {author} {\bibfnamefont {Y.}~\bibnamefont {Shiomi}}, \bibinfo {author}
  {\bibfnamefont {N.}~\bibnamefont {Nagaosa}}, \ and\ \bibinfo {author}
  {\bibfnamefont {Y.}~\bibnamefont {Tokura}},\ }\bibfield  {title} {\enquote
  {\bibinfo {title} {Observation of the magnon hall effect},}\ }\href@noop {}
  {\bibfield  {journal} {\bibinfo  {journal} {Science}\ }\textbf {\bibinfo
  {volume} {329}},\ \bibinfo {pages} {297} (\bibinfo {year}
  {2010})}\BibitemShut {NoStop}%
\bibitem [{\citenamefont {Chisnell}\ \emph {et~al.}(2015)\citenamefont
  {Chisnell}, \citenamefont {Helton}, \citenamefont {Freedman}, \citenamefont
  {Singh}, \citenamefont {Bewley}, \citenamefont {Nocera},\ and\ \citenamefont
  {Lee}}]{chisnell2015topological}%
  \BibitemOpen
  \bibfield  {author} {\bibinfo {author} {\bibfnamefont {R.}~\bibnamefont
  {Chisnell}}, \bibinfo {author} {\bibfnamefont {J.}~\bibnamefont {Helton}},
  \bibinfo {author} {\bibfnamefont {D.~E.}\ \bibnamefont {Freedman}}, \bibinfo
  {author} {\bibfnamefont {D.}~\bibnamefont {Singh}}, \bibinfo {author}
  {\bibfnamefont {R.}~\bibnamefont {Bewley}}, \bibinfo {author} {\bibfnamefont
  {D.~G.}\ \bibnamefont {Nocera}}, \ and\ \bibinfo {author} {\bibfnamefont
  {Y.~S.}\ \bibnamefont {Lee}},\ }\bibfield  {title} {\enquote {\bibinfo
  {title} {Topological magnon bands in a kagome lattice ferromagnet},}\
  }\href@noop {} {\bibfield  {journal} {\bibinfo  {journal} {Phys. Rev. Lett.}\
  }\textbf {\bibinfo {volume} {115}},\ \bibinfo {pages} {147201} (\bibinfo
  {year} {2015})}\BibitemShut {NoStop}%
\bibitem [{\citenamefont {Kondo}\ \emph {et~al.}(2019)\citenamefont {Kondo},
  \citenamefont {Akagi},\ and\ \citenamefont {Katsura}}]{kondo2019z}%
  \BibitemOpen
  \bibfield  {author} {\bibinfo {author} {\bibfnamefont {H.}~\bibnamefont
  {Kondo}}, \bibinfo {author} {\bibfnamefont {Y.}~\bibnamefont {Akagi}}, \ and\
  \bibinfo {author} {\bibfnamefont {H.}~\bibnamefont {Katsura}},\ }\bibfield
  {title} {\enquote {\bibinfo {title} {Z2 topological invariant for magnon spin
  hall systems},}\ }\href@noop {} {\bibfield  {journal} {\bibinfo  {journal}
  {Phys. Rev. B}\ }\textbf {\bibinfo {volume} {99}},\ \bibinfo {pages} {041110}
  (\bibinfo {year} {2019})}\BibitemShut {NoStop}%
\bibitem [{\citenamefont {Mook}\ \emph {et~al.}(2014)\citenamefont {Mook},
  \citenamefont {Henk},\ and\ \citenamefont {Mertig}}]{topomagnonti-1}%
  \BibitemOpen
  \bibfield  {author} {\bibinfo {author} {\bibfnamefont {A.}~\bibnamefont
  {Mook}}, \bibinfo {author} {\bibfnamefont {J.}~\bibnamefont {Henk}}, \ and\
  \bibinfo {author} {\bibfnamefont {I.}~\bibnamefont {Mertig}},\ }\bibfield
  {title} {\enquote {\bibinfo {title} {{Edge states in topological magnon
  insulators}},}\ }\href@noop {} {\bibfield  {journal} {\bibinfo  {journal}
  {Phys. Rev. B}\ }\textbf {\bibinfo {volume} {90}},\ \bibinfo {pages} {024412}
  (\bibinfo {year} {2014})}\BibitemShut {NoStop}%
\bibitem [{\citenamefont {Zhang}\ \emph {et~al.}(2013)\citenamefont {Zhang},
  \citenamefont {Ren}, \citenamefont {Wang},\ and\ \citenamefont
  {Li}}]{topomagnonti-2}%
  \BibitemOpen
  \bibfield  {author} {\bibinfo {author} {\bibfnamefont {L.}~\bibnamefont
  {Zhang}}, \bibinfo {author} {\bibfnamefont {J.}~\bibnamefont {Ren}}, \bibinfo
  {author} {\bibfnamefont {J.-S.}\ \bibnamefont {Wang}}, \ and\ \bibinfo
  {author} {\bibfnamefont {B.}~\bibnamefont {Li}},\ }\bibfield  {title}
  {\enquote {\bibinfo {title} {{Topological magnon insulator in insulating
  ferromagnet}},}\ }\href@noop {} {\bibfield  {journal} {\bibinfo  {journal}
  {Phys. Rev. B}\ }\textbf {\bibinfo {volume} {87}},\ \bibinfo {pages} {144101}
  (\bibinfo {year} {2013})}\BibitemShut {NoStop}%
\bibitem [{\citenamefont {Fransson}\ \emph {et~al.}(2016)\citenamefont
  {Fransson}, \citenamefont {Black-Schaffer},\ and\ \citenamefont
  {Balatsky}}]{topomagnondirac-1}%
  \BibitemOpen
  \bibfield  {author} {\bibinfo {author} {\bibfnamefont {J.}~\bibnamefont
  {Fransson}}, \bibinfo {author} {\bibfnamefont {A.~M.}\ \bibnamefont
  {Black-Schaffer}}, \ and\ \bibinfo {author} {\bibfnamefont {A.~V.}\
  \bibnamefont {Balatsky}},\ }\bibfield  {title} {\enquote {\bibinfo {title}
  {{Magnon dirac materials}},}\ }\href@noop {} {\bibfield  {journal} {\bibinfo
  {journal} {Phys. Rev. B}\ }\textbf {\bibinfo {volume} {94}},\ \bibinfo
  {pages} {075401} (\bibinfo {year} {2016})}\BibitemShut {NoStop}%
\bibitem [{\citenamefont {Owerre}(2017)}]{topomagnondirac-2}%
  \BibitemOpen
  \bibfield  {author} {\bibinfo {author} {\bibfnamefont {S.~A.}\ \bibnamefont
  {Owerre}},\ }\bibfield  {title} {\enquote {\bibinfo {title} {{Magnonic
  analogs of topological Dirac semimetals}},}\ }\href@noop {} {\bibfield
  {journal} {\bibinfo  {journal} {J. Phys. Commun.}\ }\textbf {\bibinfo
  {volume} {1}},\ \bibinfo {pages} {025007} (\bibinfo {year}
  {2017})}\BibitemShut {NoStop}%
\bibitem [{\citenamefont {Okuma}(2017)}]{topomagnon-new}%
  \BibitemOpen
  \bibfield  {author} {\bibinfo {author} {\bibfnamefont {N.}~\bibnamefont
  {Okuma}},\ }\bibfield  {title} {\enquote {\bibinfo {title} {{Magnon
  Spin-Momentum Locking: Various Spin Vortices and Dirac magnons in
  Noncollinear Antiferromagnets}},}\ }\href@noop {} {\bibfield  {journal}
  {\bibinfo  {journal} {Phys. Rev. Lett.}\ }\textbf {\bibinfo {volume} {119}},\
  \bibinfo {pages} {107205} (\bibinfo {year} {2017})}\BibitemShut {NoStop}%
\bibitem [{\citenamefont {Yao}\ \emph {et~al.}(2018)\citenamefont {Yao},
  \citenamefont {Li}, \citenamefont {Wang}, \citenamefont {Xue}, \citenamefont
  {Dan}, \citenamefont {Iida}, \citenamefont {Kamazawa}, \citenamefont {Li},
  \citenamefont {Fang},\ and\ \citenamefont {Li}}]{CTO-ref2}%
  \BibitemOpen
  \bibfield  {author} {\bibinfo {author} {\bibfnamefont {W.}~\bibnamefont
  {Yao}}, \bibinfo {author} {\bibfnamefont {C.}~\bibnamefont {Li}}, \bibinfo
  {author} {\bibfnamefont {L.}~\bibnamefont {Wang}}, \bibinfo {author}
  {\bibfnamefont {S.}~\bibnamefont {Xue}}, \bibinfo {author} {\bibfnamefont
  {Y.}~\bibnamefont {Dan}}, \bibinfo {author} {\bibfnamefont {K.}~\bibnamefont
  {Iida}}, \bibinfo {author} {\bibfnamefont {K.}~\bibnamefont {Kamazawa}},
  \bibinfo {author} {\bibfnamefont {K.}~\bibnamefont {Li}}, \bibinfo {author}
  {\bibfnamefont {C.}~\bibnamefont {Fang}}, \ and\ \bibinfo {author}
  {\bibfnamefont {Y.}~\bibnamefont {Li}},\ }\bibfield  {title} {\enquote
  {\bibinfo {title} {{Topological spin excitations in a three-dimensional
  antiferromagnet}},}\ }\href@noop {} {\bibfield  {journal} {\bibinfo
  {journal} {Nat. Phys.}\ }\textbf {\bibinfo {volume} {14}},\ \bibinfo {pages}
  {1011} (\bibinfo {year} {2018})}\BibitemShut {NoStop}%
\bibitem [{\citenamefont {Bao}\ \emph {et~al.}(2018)\citenamefont {Bao},
  \citenamefont {Wang}, \citenamefont {Wang}, \citenamefont {Cai},
  \citenamefont {Li}, \citenamefont {Ma}, \citenamefont {Wang}, \citenamefont
  {Ran}, \citenamefont {Dong}, \citenamefont {Abernathy}, \citenamefont {Yu},
  \citenamefont {Wan}, \citenamefont {Li},\ and\ \citenamefont
  {Wen}}]{CTO-ref3}%
  \BibitemOpen
  \bibfield  {author} {\bibinfo {author} {\bibfnamefont {S.}~\bibnamefont
  {Bao}}, \bibinfo {author} {\bibfnamefont {J.}~\bibnamefont {Wang}}, \bibinfo
  {author} {\bibfnamefont {W.}~\bibnamefont {Wang}}, \bibinfo {author}
  {\bibfnamefont {Z.}~\bibnamefont {Cai}}, \bibinfo {author} {\bibfnamefont
  {S.}~\bibnamefont {Li}}, \bibinfo {author} {\bibfnamefont {Z.}~\bibnamefont
  {Ma}}, \bibinfo {author} {\bibfnamefont {D.}~\bibnamefont {Wang}}, \bibinfo
  {author} {\bibfnamefont {K.}~\bibnamefont {Ran}}, \bibinfo {author}
  {\bibfnamefont {Z.-Y.}\ \bibnamefont {Dong}}, \bibinfo {author}
  {\bibfnamefont {D.~L.}\ \bibnamefont {Abernathy}}, \bibinfo {author}
  {\bibfnamefont {S.-L.}\ \bibnamefont {Yu}}, \bibinfo {author} {\bibfnamefont
  {X.}~\bibnamefont {Wan}}, \bibinfo {author} {\bibfnamefont {J.-X.}\
  \bibnamefont {Li}}, \ and\ \bibinfo {author} {\bibfnamefont {J.}~\bibnamefont
  {Wen}},\ }\bibfield  {title} {\enquote {\bibinfo {title} {{Discovery of
  coexisting Dirac and triply degenerate magnons in a three-dimensional
  antiferromagnet}},}\ }\href@noop {} {\bibfield  {journal} {\bibinfo
  {journal} {Nat. Commun.}\ }\textbf {\bibinfo {volume} {9}},\ \bibinfo {pages}
  {2591} (\bibinfo {year} {2018})}\BibitemShut {NoStop}%
\bibitem [{\citenamefont {Li}\ \emph {et~al.}(2016)\citenamefont {Li},
  \citenamefont {Li}, \citenamefont {Kim}, \citenamefont {Balents},
  \citenamefont {Yu},\ and\ \citenamefont {Chen}}]{topomagnon-1}%
  \BibitemOpen
  \bibfield  {author} {\bibinfo {author} {\bibfnamefont {F.-Y.}\ \bibnamefont
  {Li}}, \bibinfo {author} {\bibfnamefont {Y.-D.}\ \bibnamefont {Li}}, \bibinfo
  {author} {\bibfnamefont {Y.~B.}\ \bibnamefont {Kim}}, \bibinfo {author}
  {\bibfnamefont {L.}~\bibnamefont {Balents}}, \bibinfo {author} {\bibfnamefont
  {Y.}~\bibnamefont {Yu}}, \ and\ \bibinfo {author} {\bibfnamefont
  {G.}~\bibnamefont {Chen}},\ }\bibfield  {title} {\enquote {\bibinfo {title}
  {{Weyl magnons in breathing pyrochlore antiferromagnets}},}\ }\href@noop {}
  {\bibfield  {journal} {\bibinfo  {journal} {Nat. Commun.}\ }\textbf {\bibinfo
  {volume} {7}},\ \bibinfo {pages} {12691} (\bibinfo {year}
  {2016})}\BibitemShut {NoStop}%
\bibitem [{\citenamefont {Mook}\ \emph {et~al.}(2016)\citenamefont {Mook},
  \citenamefont {Henk},\ and\ \citenamefont {Mertig}}]{topomagnon-2}%
  \BibitemOpen
  \bibfield  {author} {\bibinfo {author} {\bibfnamefont {A.}~\bibnamefont
  {Mook}}, \bibinfo {author} {\bibfnamefont {J.}~\bibnamefont {Henk}}, \ and\
  \bibinfo {author} {\bibfnamefont {I.}~\bibnamefont {Mertig}},\ }\bibfield
  {title} {\enquote {\bibinfo {title} {{Tunable magnon Weyl points in
  ferromagnetic pyrochlores}},}\ }\href@noop {} {\bibfield  {journal} {\bibinfo
   {journal} {Phys. Rev. Lett.}\ }\textbf {\bibinfo {volume} {117}},\ \bibinfo
  {pages} {157204} (\bibinfo {year} {2016})}\BibitemShut {NoStop}%
\bibitem [{\citenamefont {Su}\ \emph {et~al.}(2017)\citenamefont {Su},
  \citenamefont {Wang},\ and\ \citenamefont {Wang}}]{topomagnon-3}%
  \BibitemOpen
  \bibfield  {author} {\bibinfo {author} {\bibfnamefont {Y.}~\bibnamefont
  {Su}}, \bibinfo {author} {\bibfnamefont {X.~S.}\ \bibnamefont {Wang}}, \ and\
  \bibinfo {author} {\bibfnamefont {X.~R.}\ \bibnamefont {Wang}},\ }\bibfield
  {title} {\enquote {\bibinfo {title} {{Magnonic Weyl semimetal and chiral
  anomaly in pyrochlore ferromagnets}},}\ }\href@noop {} {\bibfield  {journal}
  {\bibinfo  {journal} {Phys. Rev. B}\ }\textbf {\bibinfo {volume} {95}},\
  \bibinfo {pages} {224403} (\bibinfo {year} {2017})}\BibitemShut {NoStop}%
\bibitem [{\citenamefont {Serga}\ \emph {et~al.}(2010)\citenamefont {Serga},
  \citenamefont {Chumak},\ and\ \citenamefont {Hillebrands}}]{serga2010yig}%
  \BibitemOpen
  \bibfield  {author} {\bibinfo {author} {\bibfnamefont {A.}~\bibnamefont
  {Serga}}, \bibinfo {author} {\bibfnamefont {A.}~\bibnamefont {Chumak}}, \
  and\ \bibinfo {author} {\bibfnamefont {B.}~\bibnamefont {Hillebrands}},\
  }\bibfield  {title} {\enquote {\bibinfo {title} {Yig magnonics},}\
  }\href@noop {} {\bibfield  {journal} {\bibinfo  {journal} {Journal of Physics
  D: Applied Physics}\ }\textbf {\bibinfo {volume} {43}},\ \bibinfo {pages}
  {264002} (\bibinfo {year} {2010})}\BibitemShut {NoStop}%
\bibitem [{\citenamefont {Kruglyak}\ \emph {et~al.}(2010)\citenamefont
  {Kruglyak}, \citenamefont {Demokritov},\ and\ \citenamefont
  {Grundler}}]{kruglyak2010magnonics}%
  \BibitemOpen
  \bibfield  {author} {\bibinfo {author} {\bibfnamefont {V.}~\bibnamefont
  {Kruglyak}}, \bibinfo {author} {\bibfnamefont {S.}~\bibnamefont
  {Demokritov}}, \ and\ \bibinfo {author} {\bibfnamefont {D.}~\bibnamefont
  {Grundler}},\ }\bibfield  {title} {\enquote {\bibinfo {title} {Magnonics},}\
  }\href@noop {} {\bibfield  {journal} {\bibinfo  {journal} {Journal of Physics
  D: Applied Physics}\ }\textbf {\bibinfo {volume} {43}},\ \bibinfo {pages}
  {264001} (\bibinfo {year} {2010})}\BibitemShut {NoStop}%
\bibitem [{\citenamefont {Chumak}\ \emph {et~al.}(2015)\citenamefont {Chumak},
  \citenamefont {Vasyuchka}, \citenamefont {Serga},\ and\ \citenamefont
  {Hillebrands}}]{chumak2015magnon}%
  \BibitemOpen
  \bibfield  {author} {\bibinfo {author} {\bibfnamefont {A.}~\bibnamefont
  {Chumak}}, \bibinfo {author} {\bibfnamefont {V.}~\bibnamefont {Vasyuchka}},
  \bibinfo {author} {\bibfnamefont {A.}~\bibnamefont {Serga}}, \ and\ \bibinfo
  {author} {\bibfnamefont {B.}~\bibnamefont {Hillebrands}},\ }\bibfield
  {title} {\enquote {\bibinfo {title} {Magnon spintronics},}\ }\href@noop {}
  {\bibfield  {journal} {\bibinfo  {journal} {Nature Physics}\ }\textbf
  {\bibinfo {volume} {11}},\ \bibinfo {pages} {453} (\bibinfo {year}
  {2015})}\BibitemShut {NoStop}%
\bibitem [{\citenamefont {Nikitov}\ \emph {et~al.}(2015)\citenamefont
  {Nikitov}, \citenamefont {Kalyabin}, \citenamefont {Lisenkov}, \citenamefont
  {Slavin}, \citenamefont {Barabanenkov}, \citenamefont {Osokin}, \citenamefont
  {Sadovnikov}, \citenamefont {Beginin}, \citenamefont {Morozova},
  \citenamefont {Filimonov} \emph {et~al.}}]{Magnonics-4}%
  \BibitemOpen
  \bibfield  {author} {\bibinfo {author} {\bibfnamefont {S.~A.}\ \bibnamefont
  {Nikitov}}, \bibinfo {author} {\bibfnamefont {D.~V.}\ \bibnamefont
  {Kalyabin}}, \bibinfo {author} {\bibfnamefont {I.~V.}\ \bibnamefont
  {Lisenkov}}, \bibinfo {author} {\bibfnamefont {A.}~\bibnamefont {Slavin}},
  \bibinfo {author} {\bibfnamefont {Y.~N.}\ \bibnamefont {Barabanenkov}},
  \bibinfo {author} {\bibfnamefont {S.~A.}\ \bibnamefont {Osokin}}, \bibinfo
  {author} {\bibfnamefont {A.~V.}\ \bibnamefont {Sadovnikov}}, \bibinfo
  {author} {\bibfnamefont {E.~N.}\ \bibnamefont {Beginin}}, \bibinfo {author}
  {\bibfnamefont {M.~A.}\ \bibnamefont {Morozova}}, \bibinfo {author}
  {\bibfnamefont {Y.~A.}\ \bibnamefont {Filimonov}},  \emph {et~al.},\
  }\bibfield  {title} {\enquote {\bibinfo {title} {Magnonics: a new research
  area in spintronics and spin wave electronics},}\ }\href@noop {} {\bibfield
  {journal} {\bibinfo  {journal} {Phys.-Usp.}\ }\textbf {\bibinfo {volume}
  {58}},\ \bibinfo {pages} {1002} (\bibinfo {year} {2015})}\BibitemShut
  {NoStop}%
\bibitem [{\citenamefont {Lenk}\ \emph {et~al.}(2011)\citenamefont {Lenk},
  \citenamefont {Ulrichs}, \citenamefont {Garbs},\ and\ \citenamefont
  {M{\"u}nzenberg}}]{Magnonics-5}%
  \BibitemOpen
  \bibfield  {author} {\bibinfo {author} {\bibfnamefont {B.}~\bibnamefont
  {Lenk}}, \bibinfo {author} {\bibfnamefont {H.}~\bibnamefont {Ulrichs}},
  \bibinfo {author} {\bibfnamefont {F.}~\bibnamefont {Garbs}}, \ and\ \bibinfo
  {author} {\bibfnamefont {M.}~\bibnamefont {M{\"u}nzenberg}},\ }\bibfield
  {title} {\enquote {\bibinfo {title} {The building blocks of magnonics},}\
  }\href@noop {} {\bibfield  {journal} {\bibinfo  {journal} {Physics Reports}\
  }\textbf {\bibinfo {volume} {507}},\ \bibinfo {pages} {107} (\bibinfo {year}
  {2011})}\BibitemShut {NoStop}%
\bibitem [{\citenamefont {Xiang}\ \emph {et~al.}(2013)\citenamefont {Xiang},
  \citenamefont {Lee}, \citenamefont {Koo}, \citenamefont {Gong},\ and\
  \citenamefont {Whangbo}}]{xiang2013magnetic}%
  \BibitemOpen
  \bibfield  {author} {\bibinfo {author} {\bibfnamefont {H.}~\bibnamefont
  {Xiang}}, \bibinfo {author} {\bibfnamefont {C.}~\bibnamefont {Lee}}, \bibinfo
  {author} {\bibfnamefont {H.-J.}\ \bibnamefont {Koo}}, \bibinfo {author}
  {\bibfnamefont {X.}~\bibnamefont {Gong}}, \ and\ \bibinfo {author}
  {\bibfnamefont {M.-H.}\ \bibnamefont {Whangbo}},\ }\bibfield  {title}
  {\enquote {\bibinfo {title} {Magnetic properties and energy-mapping
  analysis},}\ }\href@noop {} {\bibfield  {journal} {\bibinfo  {journal}
  {Dalton Transactions}\ }\textbf {\bibinfo {volume} {42}},\ \bibinfo {pages}
  {823} (\bibinfo {year} {2013})}\BibitemShut {NoStop}%
\bibitem [{\citenamefont {Liechtenstein}\ \emph {et~al.}(1987)\citenamefont
  {Liechtenstein}, \citenamefont {Katsnelson}, \citenamefont {Antropov},\ and\
  \citenamefont {Gubanov}}]{Jref1}%
  \BibitemOpen
  \bibfield  {author} {\bibinfo {author} {\bibfnamefont {A.~I.}\ \bibnamefont
  {Liechtenstein}}, \bibinfo {author} {\bibfnamefont {M.~I.}\ \bibnamefont
  {Katsnelson}}, \bibinfo {author} {\bibfnamefont {V.~P.}\ \bibnamefont
  {Antropov}}, \ and\ \bibinfo {author} {\bibfnamefont {V.~A.}\ \bibnamefont
  {Gubanov}},\ }\bibfield  {title} {\enquote {\bibinfo {title} {{Local spin
  density functional approach to the theory of exchange interactions in
  ferromagnetic metals and alloys}},}\ }\href@noop {} {\bibfield  {journal}
  {\bibinfo  {journal} {J. Magn. Magn. Mater.}\ }\textbf {\bibinfo {volume}
  {67}},\ \bibinfo {pages} {65} (\bibinfo {year} {1987})}\BibitemShut {NoStop}%
\bibitem [{\citenamefont {Bruno}(2003)}]{Bruno}%
  \BibitemOpen
  \bibfield  {author} {\bibinfo {author} {\bibfnamefont {P.}~\bibnamefont
  {Bruno}},\ }\bibfield  {title} {\enquote {\bibinfo {title} {{Exchange
  Interaction Parameters and Adiabatic Spin-Wave Spectra of Ferromagnets: A
  ¡°Renormalized Magnetic Force Theorem¡±}},}\ }\href@noop {} {\bibfield
  {journal} {\bibinfo  {journal} {Phys. Rev. Lett.}\ }\textbf {\bibinfo
  {volume} {90}},\ \bibinfo {pages} {085205} (\bibinfo {year}
  {2003})}\BibitemShut {NoStop}%
\bibitem [{\citenamefont {Wan}\ \emph {et~al.}(2006)\citenamefont {Wan},
  \citenamefont {Yin},\ and\ \citenamefont {Savrasov}}]{ourJ1}%
  \BibitemOpen
  \bibfield  {author} {\bibinfo {author} {\bibfnamefont {X.}~\bibnamefont
  {Wan}}, \bibinfo {author} {\bibfnamefont {Q.}~\bibnamefont {Yin}}, \ and\
  \bibinfo {author} {\bibfnamefont {S.~Y.}\ \bibnamefont {Savrasov}},\
  }\bibfield  {title} {\enquote {\bibinfo {title} {{Calculation of magnetic
  exchange interactions in mott-hubbard systems}},}\ }\href@noop {} {\bibfield
  {journal} {\bibinfo  {journal} {Phys. Rev. Lett.}\ }\textbf {\bibinfo
  {volume} {97}},\ \bibinfo {pages} {266403} (\bibinfo {year}
  {2006})}\BibitemShut {NoStop}%
\bibitem [{\citenamefont {Ebert}\ \emph {et~al.}(2011)\citenamefont {Ebert},
  \citenamefont {Koedderitzsch},\ and\ \citenamefont {Minar}}]{KKR-J-1}%
  \BibitemOpen
  \bibfield  {author} {\bibinfo {author} {\bibfnamefont {H.}~\bibnamefont
  {Ebert}}, \bibinfo {author} {\bibfnamefont {D.}~\bibnamefont
  {Koedderitzsch}}, \ and\ \bibinfo {author} {\bibfnamefont {J.}~\bibnamefont
  {Minar}},\ }\bibfield  {title} {\enquote {\bibinfo {title} {Calculating
  condensed matter properties using the kkr-green's function method¡ªrecent
  developments and applications},}\ }\href@noop {} {\bibfield  {journal}
  {\bibinfo  {journal} {Rep. Prog. Phys.}\ }\textbf {\bibinfo {volume} {74}},\
  \bibinfo {pages} {096501} (\bibinfo {year} {2011})}\BibitemShut {NoStop}%
\bibitem [{\citenamefont {Secchi}\ \emph {et~al.}(2015)\citenamefont {Secchi},
  \citenamefont {Lichtenstein},\ and\ \citenamefont {Katsnelson}}]{KKR-J-2}%
  \BibitemOpen
  \bibfield  {author} {\bibinfo {author} {\bibfnamefont {A.}~\bibnamefont
  {Secchi}}, \bibinfo {author} {\bibfnamefont {A.~I.}\ \bibnamefont
  {Lichtenstein}}, \ and\ \bibinfo {author} {\bibfnamefont {M.~I.}\
  \bibnamefont {Katsnelson}},\ }\bibfield  {title} {\enquote {\bibinfo {title}
  {Magnetic interactions in strongly correlated systems: Spin and orbital
  contributions},}\ }\href@noop {} {\bibfield  {journal} {\bibinfo  {journal}
  {Annals of Physics}\ }\textbf {\bibinfo {volume} {360}},\ \bibinfo {pages}
  {61} (\bibinfo {year} {2015})}\BibitemShut {NoStop}%
\bibitem [{\citenamefont {Rosengaard}\ and\ \citenamefont
  {Johansson}(1997)}]{frozen-magnon-1}%
  \BibitemOpen
  \bibfield  {author} {\bibinfo {author} {\bibfnamefont {N.~M.}\ \bibnamefont
  {Rosengaard}}\ and\ \bibinfo {author} {\bibfnamefont {B.}~\bibnamefont
  {Johansson}},\ }\bibfield  {title} {\enquote {\bibinfo {title}
  {Finite-temperature study of itinerant ferromagnetism in fe, co, and ni},}\
  }\href@noop {} {\bibfield  {journal} {\bibinfo  {journal} {Physical Review
  B}\ }\textbf {\bibinfo {volume} {55}},\ \bibinfo {pages} {14975} (\bibinfo
  {year} {1997})}\BibitemShut {NoStop}%
\bibitem [{\citenamefont {Halilov}\ \emph {et~al.}(1998)\citenamefont
  {Halilov}, \citenamefont {Eschrig}, \citenamefont {Perlov},\ and\
  \citenamefont {Oppeneer}}]{frozen-magnon-2}%
  \BibitemOpen
  \bibfield  {author} {\bibinfo {author} {\bibfnamefont {S.}~\bibnamefont
  {Halilov}}, \bibinfo {author} {\bibfnamefont {H.}~\bibnamefont {Eschrig}},
  \bibinfo {author} {\bibfnamefont {A.}~\bibnamefont {Perlov}}, \ and\ \bibinfo
  {author} {\bibfnamefont {P.}~\bibnamefont {Oppeneer}},\ }\bibfield  {title}
  {\enquote {\bibinfo {title} {Adiabatic spin dynamics from
  spin-density-functional theory: Application to fe, co, and ni},}\ }\href@noop
  {} {\bibfield  {journal} {\bibinfo  {journal} {Physical Review B}\ }\textbf
  {\bibinfo {volume} {58}},\ \bibinfo {pages} {293} (\bibinfo {year}
  {1998})}\BibitemShut {NoStop}%
\bibitem [{\citenamefont {Paddison}(2020)}]{Paddison}%
  \BibitemOpen
  \bibfield  {author} {\bibinfo {author} {\bibfnamefont {J.~A.~M.}\
  \bibnamefont {Paddison}},\ }\bibfield  {title} {\enquote {\bibinfo {title}
  {Scattering signatures of bond-dependent magnetic interactions},}\
  }\href@noop {} {\bibfield  {journal} {\bibinfo  {journal} {Physical Review
  Lett.}\ }\textbf {\bibinfo {volume} {125}},\ \bibinfo {pages} {247202}
  (\bibinfo {year} {2020})}\BibitemShut {NoStop}%
\bibitem [{\citenamefont {Anisimov}\ \emph {et~al.}(1997)\citenamefont
  {Anisimov}, \citenamefont {Aryasetiawan},\ and\ \citenamefont
  {Lichtenstein}}]{LDA+U}%
  \BibitemOpen
  \bibfield  {author} {\bibinfo {author} {\bibfnamefont {V.}~\bibnamefont
  {Anisimov}}, \bibinfo {author} {\bibfnamefont {F.}~\bibnamefont
  {Aryasetiawan}}, \ and\ \bibinfo {author} {\bibfnamefont {A.}~\bibnamefont
  {Lichtenstein}},\ }\bibfield  {title} {\enquote {\bibinfo {title}
  {{Calculation of magnetic exchange interactions in mott-hubbard systems}},}\
  }\href@noop {} {\bibfield  {journal} {\bibinfo  {journal} {J. Phys.: Condens.
  Matter}\ }\textbf {\bibinfo {volume} {9}},\ \bibinfo {pages} {767} (\bibinfo
  {year} {1997})}\BibitemShut {NoStop}%
\bibitem [{\citenamefont {Kotliar}\ \emph {et~al.}(2006)\citenamefont
  {Kotliar}, \citenamefont {Savrasov}, \citenamefont {Haule}, \citenamefont
  {Oudovenko}, \citenamefont {Parcollet},\ and\ \citenamefont
  {Marianetti}}]{LDA+DMFT}%
  \BibitemOpen
  \bibfield  {author} {\bibinfo {author} {\bibfnamefont {G.}~\bibnamefont
  {Kotliar}}, \bibinfo {author} {\bibfnamefont {S.~Y.}\ \bibnamefont
  {Savrasov}}, \bibinfo {author} {\bibfnamefont {K.}~\bibnamefont {Haule}},
  \bibinfo {author} {\bibfnamefont {V.~S.}\ \bibnamefont {Oudovenko}}, \bibinfo
  {author} {\bibfnamefont {O.}~\bibnamefont {Parcollet}}, \ and\ \bibinfo
  {author} {\bibfnamefont {C.~A.}\ \bibnamefont {Marianetti}},\ }\bibfield
  {title} {\enquote {\bibinfo {title} {{Electronic structure calculations with
  dynamical mean-field theory}},}\ }\href@noop {} {\bibfield  {journal}
  {\bibinfo  {journal} {Rev. Mod. Phys.}\ }\textbf {\bibinfo {volume} {78}},\
  \bibinfo {pages} {865} (\bibinfo {year} {2006})}\BibitemShut {NoStop}%
\bibitem [{\citenamefont {Hohenberg}\ and\ \citenamefont
  {Brinkman}(1974)}]{sum-rule}%
  \BibitemOpen
  \bibfield  {author} {\bibinfo {author} {\bibfnamefont {P.~C.}\ \bibnamefont
  {Hohenberg}}\ and\ \bibinfo {author} {\bibfnamefont {W.~F.}\ \bibnamefont
  {Brinkman}},\ }\bibfield  {title} {\enquote {\bibinfo {title} {Sum rules for
  the frequency spectrum of linear magnetic chains},}\ }\href@noop {}
  {\bibfield  {journal} {\bibinfo  {journal} {Phys. Rev. B}\ }\textbf {\bibinfo
  {volume} {10}},\ \bibinfo {pages} {128} (\bibinfo {year} {1974})}\BibitemShut
  {NoStop}%
\bibitem [{\citenamefont {Kitaev}(2006)}]{Kitaev}%
  \BibitemOpen
  \bibfield  {author} {\bibinfo {author} {\bibfnamefont {A.}~\bibnamefont
  {Kitaev}},\ }\bibfield  {title} {\enquote {\bibinfo {title} {Anyons in an
  exactly solved model and beyond},}\ }\href@noop {} {\bibfield  {journal}
  {\bibinfo  {journal} {Annals of Physics}\ }\textbf {\bibinfo {volume}
  {321}},\ \bibinfo {pages} {2} (\bibinfo {year} {2006})}\BibitemShut {NoStop}%
\bibitem [{\citenamefont {Gardner}\ \emph {et~al.}(2010)\citenamefont
  {Gardner}, \citenamefont {Gingras},\ and\ \citenamefont
  {Greedan}}]{pyrochlore}%
  \BibitemOpen
  \bibfield  {author} {\bibinfo {author} {\bibfnamefont {J.~S.}\ \bibnamefont
  {Gardner}}, \bibinfo {author} {\bibfnamefont {M.~J.}\ \bibnamefont
  {Gingras}}, \ and\ \bibinfo {author} {\bibfnamefont {J.~E.}\ \bibnamefont
  {Greedan}},\ }\bibfield  {title} {\enquote {\bibinfo {title} {Magnetic
  pyrochlore oxides},}\ }\href@noop {} {\bibfield  {journal} {\bibinfo
  {journal} {Rev. Mod. Phys.}\ }\textbf {\bibinfo {volume} {82}},\ \bibinfo
  {pages} {53} (\bibinfo {year} {2010})}\BibitemShut {NoStop}%
\bibitem [{\citenamefont {Dzyaloshinsky}(1958)}]{Dzyaloshinsky}%
  \BibitemOpen
  \bibfield  {author} {\bibinfo {author} {\bibfnamefont {I.}~\bibnamefont
  {Dzyaloshinsky}},\ }\bibfield  {title} {\enquote {\bibinfo {title} {A
  thermodynamic theory of "weak" ferromagnetism of antiferromagnetics},}\
  }\href@noop {} {\bibfield  {journal} {\bibinfo  {journal} {J. Phys. Chem.
  Solids}\ }\textbf {\bibinfo {volume} {4}},\ \bibinfo {pages} {241} (\bibinfo
  {year} {1958})}\BibitemShut {NoStop}%
\bibitem [{\citenamefont {Moriya}(1960)}]{dm}%
  \BibitemOpen
  \bibfield  {author} {\bibinfo {author} {\bibfnamefont {T.}~\bibnamefont
  {Moriya}},\ }\bibfield  {title} {\enquote {\bibinfo {title} {Anisotropic
  superexchange interaction and weak ferromagnetism},}\ }\href@noop {}
  {\bibfield  {journal} {\bibinfo  {journal} {Physical Review}\ }\textbf
  {\bibinfo {volume} {120}},\ \bibinfo {pages} {91} (\bibinfo {year}
  {1960})}\BibitemShut {NoStop}%
\bibitem [{\citenamefont {Santini}\ \emph {et~al.}(2009)\citenamefont
  {Santini}, \citenamefont {Carretta}, \citenamefont {Amoretti}, \citenamefont
  {Caciuffo}, \citenamefont {Magnani},\ and\ \citenamefont
  {Lander}}]{multipolarnew}%
  \BibitemOpen
  \bibfield  {author} {\bibinfo {author} {\bibfnamefont {P.}~\bibnamefont
  {Santini}}, \bibinfo {author} {\bibfnamefont {S.}~\bibnamefont {Carretta}},
  \bibinfo {author} {\bibfnamefont {G.}~\bibnamefont {Amoretti}}, \bibinfo
  {author} {\bibfnamefont {R.}~\bibnamefont {Caciuffo}}, \bibinfo {author}
  {\bibfnamefont {N.}~\bibnamefont {Magnani}}, \ and\ \bibinfo {author}
  {\bibfnamefont {G.~H.}\ \bibnamefont {Lander}},\ }\bibfield  {title}
  {\enquote {\bibinfo {title} {{Multipolar interactions in f-electron systems:
  The paradigm of actinide dioxides}},}\ }\href@noop {} {\bibfield  {journal}
  {\bibinfo  {journal} {Rev. Mod. Phys.}\ }\textbf {\bibinfo {volume} {81}},\
  \bibinfo {pages} {807} (\bibinfo {year} {2009})}\BibitemShut {NoStop}%
\bibitem [{\citenamefont {Kugel}\ and\ \citenamefont
  {Khomskii}(1982)}]{KK-1982}%
  \BibitemOpen
  \bibfield  {author} {\bibinfo {author} {\bibfnamefont {K.}~\bibnamefont
  {Kugel}}\ and\ \bibinfo {author} {\bibfnamefont {D.}~\bibnamefont
  {Khomskii}},\ }\bibfield  {title} {\enquote {\bibinfo {title} {The
  jahn-teller effect and magnetism: transition metal compounds},}\ }\href@noop
  {} {\bibfield  {journal} {\bibinfo  {journal} {Sov. Phys. Usp.}\ }\textbf
  {\bibinfo {volume} {25}},\ \bibinfo {pages} {231} (\bibinfo {year}
  {1982})}\BibitemShut {NoStop}%
\end{thebibliography}%

\end{document}